\documentstyle[epsfig]{mn}
\begin{document}
\normalsize

\title[Hard X-ray states and radio emission in GRS~1915+105]
{Hard X-ray states and radio emission in GRS~1915+105}
\author[M.~Klein-Wolt et al.]
{M. Klein-Wolt$^1$\thanks{email : klein@astro.uva.nl},
 R. P. Fender$^1$, G. G. Pooley$^2$,
 T. Belloni$^3$, S. Migliari$^{1,3}$\cr
 E. H. Morgan$^4$, M. van der Klis$^1$\\
$^1$ Astronomical Institute `Anton Pannekoek', University of Amsterdam,
and Center for High Energy Astrophysics, Kruislaan 403, \\
1098 SJ, Amsterdam, The Netherlands\\
$^2$ Mullard Radio Astronomy Observatory, Cavendish Laboratory,
Madingley Road, Cambridge CB3 OHE \\
$^3$ Osservatorio Astronomico di Brera, Via. E. Bianchi 46, I-23870,
        Merate, Italy\\
$^4$ Department of Physics, Massachusetts Institute of Technology, Cambridge, MA 0213, U.S.A.\\
}

\normalsize
\maketitle

\begin{abstract}

We have compared simultaneous Ryle Telescope radio and Rossi X-Ray Timing Explorer X-ray observations of the galactic microquasar GRS~1915+105, using the classification of the X-ray behaviour in terms of three states as previously established. We find a strong (one--to--one) relation between radio oscillation events and series of spectrally hard states in the X-ray light curves, if the hard states are longer than $\sim100$s and are ``well separated'' from each other. In all other cases the source either shows a low-level or a high level radio emission, but no radio oscillation events. During intervals when the source stays in the hard spectral state for periods of days to months, the radio behaviour is quite different; during some of these intervals a quasi-continuous jet is formed with an almost flat synchrotron spectrum extending to at least the near-infrared. Based on the similarities between the oscillation profiles at different wavelengths, we suggest a scenario which can explain most of the complex X-ray~:~radio behaviour of GRS~1915+105. We compare this behaviour with that of other black hole sources and challenge previous reports of a relation between spectrally soft X-ray states and the radio emission.

\end{abstract}


\section{Introduction}

The black hole candidate GRS~1915+105 was the first Galactic source to
display apparent superluminal motions of radio-emitting ejecta
(Mirabel \& Rodr\'\i guez 1994; Fender et al. 1999b; Rodr\'\i guez \& Mirabel
1999). It is probably the best example to date of the strong coupling
between the accretion disc and the jet in a black hole system (Pooley \& Fender
1997; Eikenberry et al. 1998, 2000; Mirabel et al. 1998; Belloni, Migliari \& Fender 2000).

As an X-ray source GRS~1915+105 was discovered in 1992 with WATCH
on board GRANAT (Castro-Tirado, Brandt \& Lund 1992). Early observations, made with this instrument and with BATSE/CGRO, showed considerable X-ray variability. Since the launch of the Rossi X-ray Timing Explorer (\emph{RXTE}) in 1995 the source has been continuously monitored with the All-Sky Monitor (ASM), and frequently observed with the Proportional Counter Array (PCA) and the High Energy X-ray Timing Experiment (HEXTE). From these observations the
extraordinary X-ray variability displayed by GRS~1915+105 became clear
(see for instance Greiner, Morgan \& Remillard 1996; Belloni et al. 1997a,b; Muno, Morgan \& Remillard 1999; Belloni et al. 2000a). Belloni et al. (2000a) found that the entire variability of the source is made up of a limited number of distinct variability classes; they identified twelve classes that each were observed to recur almost identically after intervals of months to even years. This behaviour is unique to GRS~1915+105, and is not observed in any other X-ray binary known to date.

As in most black hole sources the X-ray spectrum of GRS~1915+105 can be modelled as a combination of two components: a soft (kT$\sim$1 -- 2 keV) disc black body and a hard power law extending to $\geq$100 keV, generally interpreted as representing the radiation from the accretion disc and a Comptonizing region (`corona'), respectively. Belloni et al. (2000a) found that the complex X-ray variability of GRS~1915+105 can be reduced to the transitions between three basic spectral states which they called A, B and C. The
spectrally soft states A and B correspond to an observable inner
accretion disc with different temperatures: in State B the inner
temperature ($\sim2.2$ keV) is higher compared to State A ($\sim1.8$ keV). For the spectrally hard State C the inner part of the accretion disc is either missing or just unobservable, and the 2 -- 25 keV spectrum is dominated by a power law of spectral index $\Gamma\sim$1.3 -- 2.4. The description of the behaviour of GRS~1915+105 as transitions between these three spectral states differs from one based on the canonical black hole states (for a review on the canonical black hole states, see for instance Tanaka \& Lewin 1995 and van der Klis 1995), as it is based on the positions in the X-ray colour-colour diagram, and not on the combined spectral and timing characteristics. There are, however, indications that the States A, B and C are related to the High State (HS), Very High state (VHS) and the Low State (LS) respectively (Belloni et al. 2000a and references therein). The description of the complex X-ray behaviour of GRS~1915+105 as the transitions between three basic spectral states, and the
classification of the light curves into twelve different variability
classes will be discussed in some more detail below.

GRS~1915+105 is also amongst the most spectacular radio sources in our
Galaxy, displaying apparent superluminal motions (Mirabel \& Rodr\'\i guez
1994; Fender et al. 1999b; Rodr\'\i guez \& Mirabel 1999) and
radio oscillations which are clearly related to quasi-periodic dips in
the X-ray light curve on time-scales of tens of minutes
(Pooley \& Fender 1997; Mirabel et al. 1998). The synchrotron
emission of these oscillations extends well beyond the radio to the
millimeter and near-infrared regimes (Fender et al. 1997; Eikenberry
et al. 1998, 2000; Mirabel et al. 1998; Fender \& Pooley 1998, 2000).
The power required to generate these repeated synchrotron
events is likely to be a very significant fraction of the entire
accretion energy of the system (Fender \& Pooley 2000, Meier 2001). There is
evidence that the power into the jet (assuming that a jet is the
origin of the radio oscillations) may be anticorrelated with the accretion
rate as inferred from X-ray spectral fits (Belloni et al. 2000b).

Belloni et al. (1997a,b) suggested that the X-ray dips correspond to
the disappearance of the inner ($\leq 100$ km) of the accretion
disc. The relation of the radio oscillations to these dips led to the
assertion that at least some of the inner disc which had `disappeared'
had in fact been accelerated and ejected from the system, and had not
simply fallen into the black hole. Based on the similarity between oscillation decays at different wavelengths, Fender et al. (1997) came to the conclusion that material is being ejected in the form of jets where the synchrotron emitting ejecta are predominantly subjected to adiabatic expansion losses. Whatever the interpretation, the coupling between radio--mm--infrared emission which is believed to arise in the jet, and the X-ray emission from the disc and corona, is very clear, and GRS~1915+105 is our most accessible route to date to an understanding of the connection between accretion and ejection around a black hole.

In this paper we cross-correlate the entire database of radio
observations with the Ryle Telescope (RT) at 15 GHz over the period
1996 -- 1999, with pointed X-ray observations with the \emph{RXTE}. In addition we use some data from the Green Bank Interferometer (GBI) variable source monitoring program. The next two sections deal with the observation analysis and the X-ray~:~radio overlap. In Section~\ref{sec:obs} we present our results by discussing a selection of representative observations in detail. In Sections~\ref{sec:emp} and ~\ref{sec:platstates} the empirical characteristics of the X-ray~:~radio correlation are discussed, and Section~\ref{sec:delays} deals with the profiles of the radio oscillation events. In Section~\ref{sec:discussion} we discus a possible scenario which explains the correlated X-ray~:~radio behaviour, and finally in Section~\ref{sec:conclusions} we present our conclusions.

\section{Observations}

\subsection{\emph{RXTE} observations}

\subsubsection{Light curve and color-color diagram extraction}
\label{sec:lcandccd}

We present 101 X-ray Proportional Counter Array (PCA) observations which overlap with the Ryle Telescope (RT) radio observations, made between 1996 June 12 and 1999 December 13 (see Table~\ref{t1}). Each observation consists of one or more intervals (observation intervals) which are separated by earth occultations. Following Belloni et al. (2000a), for each observation interval we produced three light curves (1 second time resolution), one in each of the following PCA channel bands: \emph{a}: 0 -- 13 (2 -- 5 keV), \emph{b}: 14 -- 35 (5 -- 13 keV) and \emph{c}: 36 -- 255 (13 -- 60 keV). From these light curves we calculate two X-ray colours:
\begin{equation}
HR_{1}=\frac{\emph{b}}{\emph{a}}
\end{equation}
\begin{equation}
HR_{2}=\frac{\emph{c}}{\emph{a}},
\end{equation}
and the total count rate (=\emph{a+b+c}). The two colours are defined
in this way to ensure linearity: any linear combination of two
spectral models lies on a straight line connecting their locations in
the colour-colour diagram (CD; HR$_{1}$ vs. HR$_{2}$). For the background we subtracted a constant level of 10, 20, and 100 counts in the
\emph{a}, \emph{b} and \emph{c} bands respectively, as determined from the analysis of typical background spectra. Notice that because 
our data set consists of observations from different \emph{RXTE} gain epochs, we have
to consider the gain changes. Changes in the PCA gain will have an effect
on the two X-ray colours resulting in an overall shift in the CD. Therefore, gain changes prevent a direct comparison between colours from different epochs. However, as the gain changes occur on timescales much longer than the typical timescales of one observation, we assume here the influence on the X-ray colours is negligible.

\subsubsection{Describing the X-ray variability}
\label{sec:xrayvar}

Following Belloni et al. (2000a) we analyse the 101 X-ray observations overlapping with our radio data set, and find that most can be assigned to one of their twelve different variability classes (see Table~\ref{allcl}). This is shown in Table~\ref{t1} where we give for each observation the MJD, the \emph{RXTE} observation ID-number, the X-ray~:~radio overlap time in minutes, the X-ray class, the length of the short and long State C intervals, and a summary of the radio behaviour\footnote[1]{Observations from \emph{RXTE} gain epochs 1 and 2 belonging to class $\chi$ were subdivided into four subclasses $\chi_{1}$, $\chi_{2}$, $\chi_{3}$ and $\chi_{4}$ based on the position in the ASM light curve (see Belloni et al. 2000a). This subdivision was not made for observations from epochs 3 and 4.}. State C intervals are referred to as long or short depending on whether they are longer or respectively shorter than 100s. The distinction between different classes of variability is made based on the light curve and the corresponding CDs: in each variability class the pattern in the X-ray light curves and in the CDs are the same, while observations from different classes show a different pattern (see also figure 2 in Belloni et al. 2000a).

\begin{figure}
\centerline{\psfig{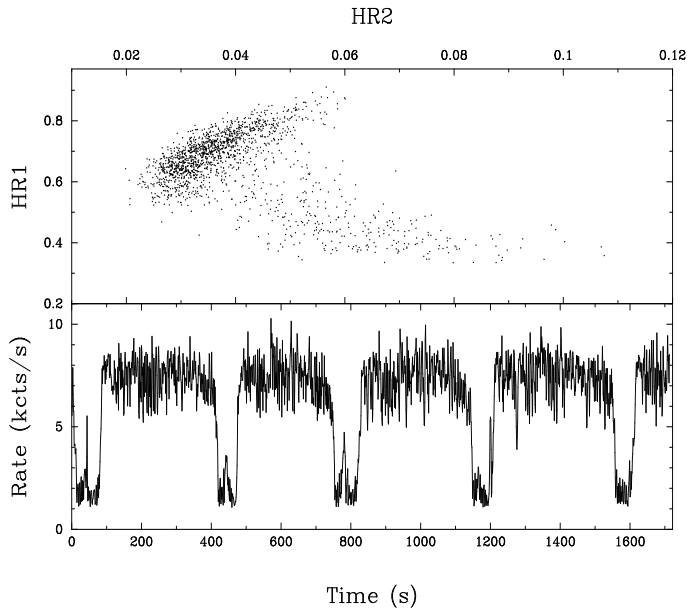}{\hfil}}
\caption[]{A representative observation in epoch 4 belonging to class $\omega$ (observation ID is P-27-00 ($\#2$)). In the top panel we show the CD and in lower panel the light curve. The colours are defined as explained in section~\ref{sec:lcandccd}. The dips, during which the source moves to the lower right in the CD, correspond to state C. Outside the dips the source makes rapid transitions between the states A and B, reminiscent of the A -- B transitions observed in class $\gamma$ observations (see figure 2 Belloni et al. 2000a). As in all the variability classes defined by Belloni et al. (2000a) the source makes the transitions in the order A -- B -- C -- A; there are no direct transitions from state C to B, but the source always first goes through state A (Belloni et al. 2000a). \label{fig:omega}}
\end{figure}

Most of the observations from our data sample could be classified as
described above. There were, however, twelve observations (see Table~\ref{t1}, and Fig.~\ref{fig:omega}) which showed X-ray behaviour that could not be classified into one of the known twelve variability classes. Because all the light curves and the CDs showed a similar pattern, that did not show any resemblance to previously classified observations (compare Fig.~\ref{fig:omega} with figure 2 in Belloni et al. 2000a) we assigned these observations to an additional variability class: class $\omega$. As already stated by Belloni et al. (2000a), the classification of all the observations into variability classes is at this stage mostly a tool, allowing one to group similar observations together and thereby making the complex behaviour of GRS~1915+105 manageable.

The description of the observations, in terms of the basic spectral States A, B and C, is now a relatively straightforward exercise as the decomposition of observation intervals in each of the twelve variability classes into the three states was already given by Belloni et al. (2000a). Furthermore we can make use of the characteristics of the three spectral states as given by Belloni et al. (2000a):
\begin{itemize}
\item{State C: low count rate, hard location in the CD (HR$_{1}$$\sim$1, HR$_{2}$$>$0.1);}
\item{State A: sharp dip in the light curve, soft location in the CD (HR$_{1}$$<$1.1, HR$_{2}$$<$0.1 );}
\item{State B: high count rate, soft location in the CD (HR$_{1}$$>$1, HR$_{2}$$<$0.1).}
\end{itemize}
Of course, due to the gain changes, the numbers given are only valid
for gain epochs 1 and 2. The observations from epoch 3 used in this analysis consisted only of class $\chi$ and $\rho$ (see Table~\ref{t1}), and the states could very easily be identified from the light curves. In the case of epoch 4 the three states still correspond to distinct positions in the CD (although
these positions are different compared to epochs 1, 2 and 3), and we found: State A: HR$_{1}$$<$0.5, HR$_{2}$$<$0.06; State B: HR$_{1}$$>$0.5, HR$_{2}$$<$0.06; State C: HR$_{1}$$\sim$0.4, HR$_{2}$$>$0.04.

\begin{figure*}
\centering{\epsfig{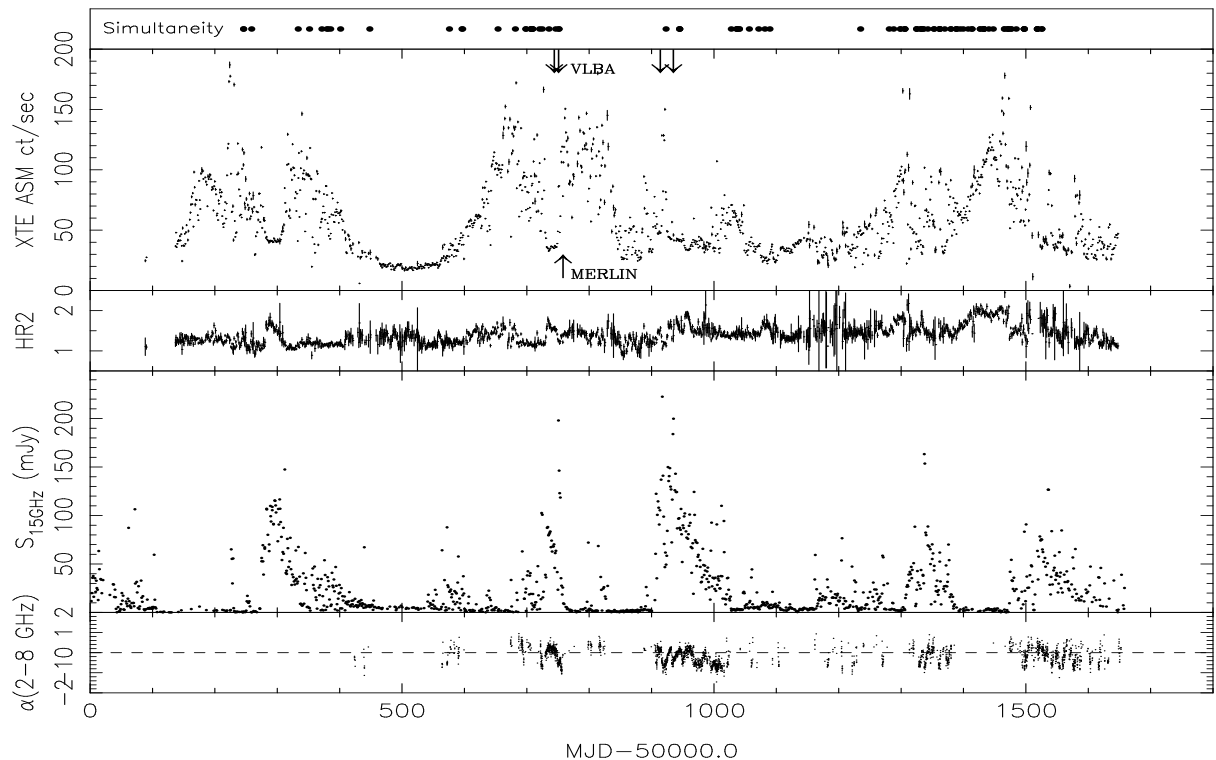}}
\caption[]{The $\sim 4$-year radio and X-ray flux history of GRS
1915+105, using daily-averaged data. The top panel indicates the
periods of overlapping RT and \emph{RXTE} PCA observations which are presented
in this paper. The other panels show the total \emph{RXTE} ASM count rate in the 2 -- 12 keV band, the ASM HR2 colour (5 -- 12/3 -- 5 keV), radio flux density at 15 GHz as measured with the RT, and the 2.3 -- 8.3 GHz radio spectral index as measured by the GBI. For the spectral index we only show detections at more than 5$\sigma$ at both GBI wavelengths. Also indicated are the epochs of the MERLIN observations of Fender et al. (1999b) and the VLBA observations of Dhawan et al. (2000).}
\label{summary2000}
\end{figure*}

\subsubsection{X-ray energy spectra}
\label{sec:xrayspectra}

We extracted energy spectra with an integration time of 16~s from
{\tt Standard2} data in the energy range 3 -- 25 keV (all Proportional Counter Units where used when possible, and all layers where added). From each spectrum, we subtracted the background estimated with {\tt Pcabackest}~2.1e.
Detector response matrices were produced using {\tt Pcarsp}~2.43, and no correction for the deadtime was made, as we expect this effect to be negligible for the purpose of this analysis. We performed fits using the `standard' model for black hole candidates, consisting of the superposition of a multi-temperature disc blackbody and a powerlaw component (see Tanaka \& Lewin 1995). Correction for interstellar absorption and an additional Gaussian emission line to take into account an excess around 6.4 keV were also included. A systematic error of 1 per cent was added. 

Belloni et al (2000a) found evidence for a variable N$_{\rm{H}}$ during the class $\chi$ intervals (State C only, see Belloni et al. 2000a) in GRS~1915+105. Since leaving the interstellar column density N$_{\rm{H}}$ free to vary during the fits would affect the values of the inner radius and make a comparison between observations more complicated, we decided to fix it. From an analysis of class $\chi$ observations we found an best fit value of 7$\times$10$^{22}$~cm$^{-2}$.
 
For the observations presented Section~\ref{sec:obs}, we isolated and analysed only the hard, low-flux intervals corresponding to State C (see Belloni et al. 2000a). This is because, with 16~s resolution, we are not able to track the fast variations in the remaining parts of the observations. For observation K-42-00 there is no State C, but we were able to analyse the whole observation. This is also the case for observation M-02-00, which consists of State A and C. The observations P-30-00/01/02/03 does not consist of any State C, but have rapid State A dips. Here again, the fast variations prevent reasonable spectral fitting. 

Finally, observation P-16-00/01/02 consists of State C only. In this case the energy spectrum is dominated by a hard power law component, and the soft (multi-temperature) disc black body component is not well constrained during the 16s spectral fits. This prevented reasonable spectral fitting. However, in order to get an estimate of the inner radius we performed fits of each of the three individual observation intervals by adding the data of each of those intervals together.

From the spectral fits, performed as explained above, the inner radius of the accretion disc is derived from the disc blackbody normalization assuming a distance of 11~kpc and a disc inclination angle of 66$^\circ$ to the line of sight (Fender et al. 1999b). The results are presented in Section~\ref{sec:obs}. Note that values of the inner radius are most likely underestimated (see Merloni, Fabian, Ross 2000) and may also depend on the choice of N$_{\rm{H}}$ (as explained above). Therefore, the values of the inner radius are only used here as an indication.

\subsection{Ryle Telescope observations}

GRS~1915+105 is monitored regularly at 15 GHz with the Ryle Telescope. In bright (more than a few mJy) states data can be obtained with up to 8-second time resolution. In this paper we use one-day averaged (Fig.~\ref{summary2000}) and 32-sec radio data (Fig.~\ref{abc50381}). The rms noise on a single 32-sec sample is typically 5 -- 6 mJy. For further details see Pooley \& Fender (1997).

\subsection{Green Bank Interferometer observations}

The Green Bank Interferometer (GBI) monitored variable radio sources
approximately daily simultaneously at $2.3$ and $8.3$ GHz. During periods
when GRS~1915+105 is bright ($\ga$~$10$ mJy) it is able to provide high ($>5\sigma$) signal-to-noise spectral information, which is utilized in Fig.~\ref{summary2000}. For more details see e.g. Waltman et al. (1994).

\normalsize
\begin{table*}
\caption{List of the 101 simultaneous \emph{RXTE} PCA -- RT observations of GRS~1915+105 up to Dec 31, 1999. The letters I,J,K,L,M,N,O,P,Q and R stand for 10408-01, 10258-01, 20402-01, 20187-02, 20186-03, 30402-01, 30703-01, 40703-01, 40403-01, and 40115-01 respectively. We give: the observing date (in MJD), the \emph{RXTE} PCA observation ID, the X-ray variability class as defined by Belloni et al. (2000a) a summary of the radio behaviour, and the length of the State C interval (in sec). In column 2 and 4 the ``/'' is used to make a distinction between a sequence of observation IDs. In column 4 a ``,'' is used to make a distinction between several observation intervals (also referred to as ``orbits'') for the same observation ID. The radio fluxes, given in column 5, are measured at 15 GHz and when present we give the Period (P) of the radio oscillation events. In the last column we define a long State C as an interval during which, on a time-scale of $\sim$100 seconds, no fast transitions are observed to other states (A and B). The remark ``entirely C'' refers to observations which consist of State C only and which are always longer than 100 seconds. The symbol ``--'' was used when there are no long and/or short State Cs. The observations presented in Fig.~\ref{abc50381} are shown in bold. Note that due to the delay between the radio and the X-ray emission, and the relatively short overlaps between the two bands, it is possible that not all the long State C intervals can be associated to radio oscillation events just by looking at the table.\label{t1}}

\begin{tabular}{|ccc|c|c|cc|}
\hline
Date   & \emph{RXTE} PCA & Overlap & X-ray & Radio behaviour	&\multicolumn{2}{|c}{Length C intervals (s)} \\
(MJD)  & OBS-ID  & (min)   & class & at 15 GHz (RT)	&Long 	&Short  \\
\hline
50246 & I-14-01/02 & 37 & $\delta$ & Undetected ($<1.2$ mJy) &-- &-- \\
50259 &  I-17-03  & 109 & $\delta$   & Undetected ($<0.9$ mJy)	&-- &--\\
50333 &  I-33-00  & 305 & $\chi_4$    & Rapid decay 120 $\rightarrow$ 20 mJy	&entirely C &--\\ 
50351 &  J-10-00 & 148 &  $\mu$         & Weak ($4.6 \pm 0.4$ mJy) 	&-- &34\\
50371 & I-41-00 & 314 &  $\nu$         & Weak ($2.8 \pm 0.3$ mJy)	&715 &$\sim$100\\
50379 &  I-43-00 & 156 & $\chi_4$ & Weak ($5.8 \pm 0.4$ mJy) 	&entirely C	&-- \\
\textbf{50381} &  \textbf{I-44-00} & 206 & $\nu$         & Oscillations ($P \sim 17$ min, $\Delta S \sim 60$ mJy)	&1100--1900 &$\sim$100\\
50385 &  I-45-00 & 50 & $\chi_4$ & Weak ($6.7 \pm 0.4$ mJy) 	&entirely C &-- \\
50401 &  K-02-02 & 154 & $\chi_4$, $\nu$~$^{a}$  & Weak rise 5 $\rightarrow$ 30 mJy &220 &--\\
50448 &  K-09-00 & 59 & $\chi_2$   & Weak ($5.6 \pm 0.4$ mJy) &entirely C &-- \\
50576 &  L-01-00 & 37 & $\alpha$          & Slow modulation 10--30 mJy &350--2500 &$\sim$80\\
50596 &   K-30-01& 50 & $\chi_4$, $\alpha$~$^{b}$   & Slow modulation 10--50 mJy &1950 ($\chi$),600 ($\alpha$) &$\sim$50 \\
50597 &  K-30-02 & 2  & $\alpha$  & Noisy \& variable : 5--35 mJy &492 &$\sim$80\\
50653 &  K-39-00 & 40 & $\gamma$  & Weak ($1.7 \pm 0.5$) &-- &-- \\
\textbf{50681} &  \textbf{K-42-00} & 26 & $\delta$  & Undetected ($<1.2$ mJy) &-- &-- \\
\textbf{50698} &  \textbf{K-45-00/01} & 440& $\beta$  & Erratic oscillations ($P \sim 40$, $\Delta S \sim 70$ mJy) &90--540 &$\sim$20 -- 50\\
\textbf{50706} & \textbf{M-02-00}/01  & 540& $\theta$          & Weakly variable 5--30 mJy & 450--750 &--\\
50707 &  M-02-03/04 & 526& $\theta$          & Weak decay 15 $\rightarrow$ 5 mJy &630--4650 &--\\
50708 &  M-02-05 & 563 &   $\chi$       & Weak ($3.2 \pm 0.2$ mJy) &entirely C &-- \\
50709 &  M-02-06 & 101 &  $\chi$   & Single 50 mJy flare &entirely C	&-- \\
50720 &  K-48-00 & 127 &  $\chi_4$  & Weak ($3.8 \pm 0.6$ mJy) &entirely C &-- \\
50724 &  L-03-00 & 39 &   $\chi$           & Bright $\sim 100$ mJy &entirely C	&-- \\
50746 &  K-52-00 & 240 &      $\chi_3$       & Steady at 60--70 mJy &entirely C &-- \\
50751 &  K-52-01/02 & 68 &   $\beta$        & Oscillations ($P \sim 22$,  $\Delta S \sim 90$ mJy) &535--758 &$\sim$40--100\\
50923 &  N-11-00 & 141 &     $\chi$        & Variable $\sim 50$ mJy &entirely C &-- \\
50945 &  N-12-01/02/03 & 32 & $\chi$             & Variable $\sim 50$ mJy &entirely C  &--  \\
51027 &  O-27-00 & 49 &   $\rho$  & Weak ($3.0 \pm 0.3$ mJy) &-- &$\sim$80\\
51036 &  O-28-00/01 & 159 &  $\rho$ & Weak ($3.3 \pm 0.3$ mJy) &-- &$\sim$80\\
51040 &  O-29-00 & 43 &  $\rho$ & Weak ($3.4 \pm 0.2$ mJy) &-- &$\sim$80\\
51056 &  O-31-00 & 26 &  $\chi$ & Weak ($10.4 \pm 0.3$ mJy) &entirely C  &--  \\
51071 &  O-33-00 & 75 &  $\chi$ & Weak ($5.0 \pm 0.6$ mJy) &entirely C  &--  \\
51081 &  O-35-00 & 128 & $\chi$ & Weak ($7.4 \pm 0.4$ mJy) &entirely C  &-- \\ 
51089 &  O-36-00 & 65 &  $\chi$ & Weak ($7.6 \pm 0.4$ mJy) &entirely C  &-- \\
--\emph{table continues}-- & & & & & & \\
\hline
\end{tabular}
\medskip
\footnotesize
\newline 
\raggedright
\emph{$^{(a)}$Orbit numbers 1,2 and 3 belong to class $\chi_{4}$, and orbit number 4 belongs to class $\nu$.}\\
\emph{$^{(b)}$Orbit number 1 belongs to class $\chi$ and orbit number 2 to class $\alpha$.}
\end{table*}
\begin{table*}
\contcaption{}
\begin{tabular}{|ccc|c|c|cc|}
\hline
Date   & \emph{RXTE} PCA & Overlap & X-ray & Radio behaviour	&\multicolumn{2}{|c}{Length C intervals (s)}  \\
(MJD)  & OBS-ID  & (min)   & class & at 15 GHz (RT)	&Long 	&Short  \\
\hline
51235 &  P-07-00 & 13 &  $\rho$ & Weak ($4.9 \pm 0.5$ mJy) &-- &$\sim$80\\
51281 &  Q-05-02 & 16 &  $\rho$ & Weak ($3.0 \pm 0.2$ mJy) &-- &$\sim$80\\
51288 &  P-12-00 & 157 &  $\kappa$  & Weak ($3.0 \pm 0.5$ mJy) &-- &$\sim$20--40\\
51299 & P-13-00/01 & 191 &  $\gamma$/$\omega$ & Weak ($2.1 \pm 0.3$ mJy) &-- &$\sim$100 \\
51306 & P-14-00/01/02 & 132 & $\kappa$  & Weak ($2.6 \pm 0.3$ mJy) &-- &$\sim$20\\
51325 &  P-15-00/01/02 & 133 &    $\chi$   & Steady decay 30 $\rightarrow$ 15 mJy &entirely C  &-- \\
\textbf{51332} &  \textbf{P-16-00/01/02} & 78 & $\chi$       & Steady $\sim 45$ mJy &entirely C  &--  \\
51335 &  P-17-00/01 & 170 & $\chi$         & Steady decay 35 $\rightarrow$ 10 mJy &entirely C  &--  \\
\textbf{51342} &  \textbf{P-18-00} & 141 &   $\beta$          & Oscillations ($P \sim 29$, $\Delta S \sim 100$ mJy) &130--993 &$\sim$40--100\\
51352 &  P-19-00/01 & 113 & $\beta$         & Irregular oscillations 0--70 mJy &1120--1460 &--\\
51361 & P-20-00/01/02/03 & 211&  $\chi$    & Decay 70 $\rightarrow$ 40 mJy &entirely C  &--\\
51370 &  P-21-01 & 118 &   $\chi$          & Steady $\sim$10--30 mJy &entirely C  &--  \\
51379 &  P-22-00/01 & 52 &  $\beta$         & Irregular oscillations 0--70 mJy &404--1400 &$\sim$40--150\\ 
51388 &   P-23-00/01/02 & 243 &  $\rho$    & Weak ($5.1 \pm 0.3$ mJy) &-- &$\sim$80\\
51394 &  P-24-00 & 241 & $\kappa$     & Weak ($3.3 \pm 0.4$ mJy) &-- &$\sim$10-90\\
51399 &  P-25-00 & 228 &   $\kappa$   & Weak ($3.3 \pm 0.7$ mJy) &-- &$\sim$10-90\\
51407 & P-26-00 & 227 & $\kappa$ & Weak ($2.5 \pm 0.6$ mJy) &-- &$\sim$10-90\\

51413 & P-27-00 & 303 & $\omega$/$\gamma$,$\omega$~$^{c}$  & Weak ($3.0 \pm 0.2$ mJy) &-- &$\sim$100 \\

51426 & P-29-01 & 95 & $\omega$/$\gamma$,$\omega$~$^{d}$ & Weak ($1.7 \pm 0.4$ mJy) &-- &$\sim$100 \\

\textbf{51432} & \textbf{P-30-00/01/02/03} & 118 & $\kappa$/$\gamma$~$^{e}$ & Weak ($1.8 \pm 0.3$ mJy) &-- &$\sim$20-40\\
51440 & R-07-00 & 53  & $\gamma$ & Weak ($4.1 \pm 0.5$ mJy) &-- &-- \\
51447 & P-31-00 & 244 & $\gamma$  & Weak ($1.6 \pm 0.3$ mJy) &-- &-- \\
51465 & P-33-01/02 &125 & $\delta$  & Weak ($3.2 \pm 0.4$ mJy) &-- &-- \\
51471 & P-34-00/01/02 &51 & $\delta$ & Undetected ($<1.1$ mJy) &-- &-- \\
51476 & P-35-00/01 &197& $\beta$ &  Oscillations ($P \sim 22$, $\Delta S \sim 60$ mJy) &200--1010 &--\\
51484 & P-36-02 & 84 & $\theta$ & Variable 15-40 mJy &192--640 &--\\
51497 & P-38-00/01/02/03 & 151 & $\theta$/$\chi$~$^{f}$ & Steady decay 140 $\rightarrow$ 40 mJy &411--1920 &--\\
51518 & P-41-00/01/02/03 & 117 & $\chi$ & Variable 35-85 mJy &entirely C  &-- \\
51525 & P-42-01/03 & 56 & $\chi$ & Variable 55-105 mJy &entirely C  &-- \\ 
\hline
\end{tabular}
\medskip
\footnotesize
\newline
\raggedright
\emph{$^{(c)}$P-27-00 is classified as class $\omega$, and for P-27-01 orbit number 1 is classified as class $\gamma$ and orbit number 2 as class $\omega$.}\\
\emph{$^{(d)}$P-29-01 is classified as class $\omega$, and for P-29-02 orbit 1 is classified as class $\gamma$ and orbit 2 as class $\omega$.}\\
\emph{$^{(e)}$All the orbits belong to class $\kappa$, except orbit numbers 2 and 3 from P-30-03 these belong to class $\gamma$.}\\
\emph{$^{(f)}$P-38-00/03 belong to class $\theta$ and P-38-01/02 belong to class $\chi$.}\\
\end{table*}

\section{Correlating observations : the overall picture}
\label{sec:overall}
We have cross-correlated the RT and \emph{RXTE} PCA observing logs to find
simultaneous data, for all observations up to 1999 Dec 31. We found
101 overlapping observations, totalling 9170 minutes (550 ksec; Table~\ref{t1}). It is clear that the density of overlapping observations became greater more recently as coordination between RT and \emph{RXTE} improved.

Before presenting individual high time-resolution observations, it is
worth looking at the remarkable overall pattern of radio and X-ray
emission from GRS~1915+105 over the past $\sim 4$ years. This is
presented in Fig.~\ref{summary2000}, where soft X-ray monitoring with
the \emph{RXTE} ASM, the ASM HR2 X-ray hardness ratio, RT radio monitoring at
15 GHz, and 2.3 -- 8.3 GHz radio spectral index (from GBI) are
plotted. There are clearly 6 or 7 major radio flaring events with flares up to a few hundred mJy. Most of these radio flares seem to be related to a low ASM count rate interval, in combination with a high and reasonable stable ASM HR2 level. However, not every low ASM count rate, high ASM HR2 interval can be related to a radio flare. During the radio flares the radio spectral index ($\alpha$, where $S_{\nu} \propto \nu^{\alpha}$) becomes negative, typically $\leq -0.5$, so that flux densities at lower frequencies reach even higher values (up to $\sim 1$ Jy at 843 MHz). Higher time resolution radio observations are known to reveal a wealth of radio behaviour not apparent in Fig.~\ref{summary2000} (e.g. Pooley \& Fender 1997).

\begin{table*}
\caption{For each variability class, we give the average maximum radio flux an indication of the radio behaviour. The terms ``weak'', ``variable'' and ``steady'' refer to the radio flux level, and the symbol ``--'' is used when there was no overlap between the radio and the X-rays. For each variability class we also give and their constituent states (Belloni et al. 2000a) and the number of times it occured in our data set.\label{allcl}}
\begin{tabular}{ccccc}
\hline
X-ray & constituent& Occurence & Average radio flux & Radio behaviour\\
class &   states   &   & at 15 GHz (mJy) & At 15 GHz \\
\hline
$\alpha$ & A B C & 3 & $\sim40$& variable but no oscillations \\
$\beta$ & A B C & 6 & $\sim80$& strong oscillations \\
$\gamma$ & A B & 7 & $\sim2.0$ & weak, no oscillations \\
$\delta$ & A B & 5 & $\la1.5$ & very weak, no oscillations \\
$\theta$ & A C & 4 & $\sim60$ & weak or variable, no oscillations \\
$\kappa$ & A B C & 5 & $\sim3.0$ & weak, no oscillations \\
$\lambda$ & A B C & 0 & -- & --\\
$\mu$ & A B C & 1 & $\sim5.0$ & weak no oscillations \\
$\nu$ & A B C & 2 & $\sim30$ &weak or strong oscillations \\
$\rho$ & A B C & 6 & $\sim4.0$ & weak, no oscillations\\
$\phi$ & A & 0 & -- & --\\
$\chi$ & C & 25 & $\sim5$--$40$ & steady, no oscillations\\
$\omega$ & A B C & 5 & $\sim2.0$ & weak, no oscillations\\
\hline
\end{tabular}
\end{table*}

\section{Simultaneous \emph{RXTE} PCA and RT observations}
\label{sec:obs}

As already mentioned, Table~\ref{t1} summarizes all the overlapping \emph{RXTE} PCA and RT observations of GRS~1915+105 up to 1999 Dec 31. In Table~\ref{allcl} we group these into the 13 X-ray `classes' (consisting of the 12 previously known classes and the new class defined in Section~\ref{sec:xrayvar}), and their three constituent `states' (Belloni et al. 2000a). Careful examination of Table~\ref{allcl} reveals that we encounter four different types of radio behaviour related to the X-ray classes. First of all there are times when the source shows a low level radio flux (designated as ``weak'' with a typical level of a few mJy; classes $\gamma$, $\delta$, $\kappa$, $\mu$, $\rho$ and $\omega$) with no radio oscillations. Secondly, there are observations which have a high radio flux level (typically between the 30 -- 80 mJy) and show radio oscillations (classes $\beta$ and $\nu$). Thirdly, we have a large number of observations (classified as class $\chi$) of which about $50$\% have a high level of radio flux ($\sim40$ mJy) while in the other $50$\% it is much lower ($\la10$mJy), but show no radio oscillations. And finally there are observations of class $\alpha$ and $\theta$ which show a relatively high level of radio emission ($\sim40$ -- $60$mJy) with significant variations, but no radio oscillation events. At this point in our argument it is still unclear which spectral state (A, B or C) is related to the radio emission, however, the only times (4 out of 101 observations) when the radio emission was undetectable were in class $\delta$ which consist of States A and B only. This suggests that State C might be relevant for the production of the (high flux) radio oscillation events.

We now describe in detail several of the high time-resolution overlapping observations, in order to clarify this behaviour. The observations discussed below are a representative subset chosen to highlight the key points of the X-ray~:~radio relation. 

\subsection{MJD50381 (1996 October 25)}

\begin{figure*}
\centering{\epsfig{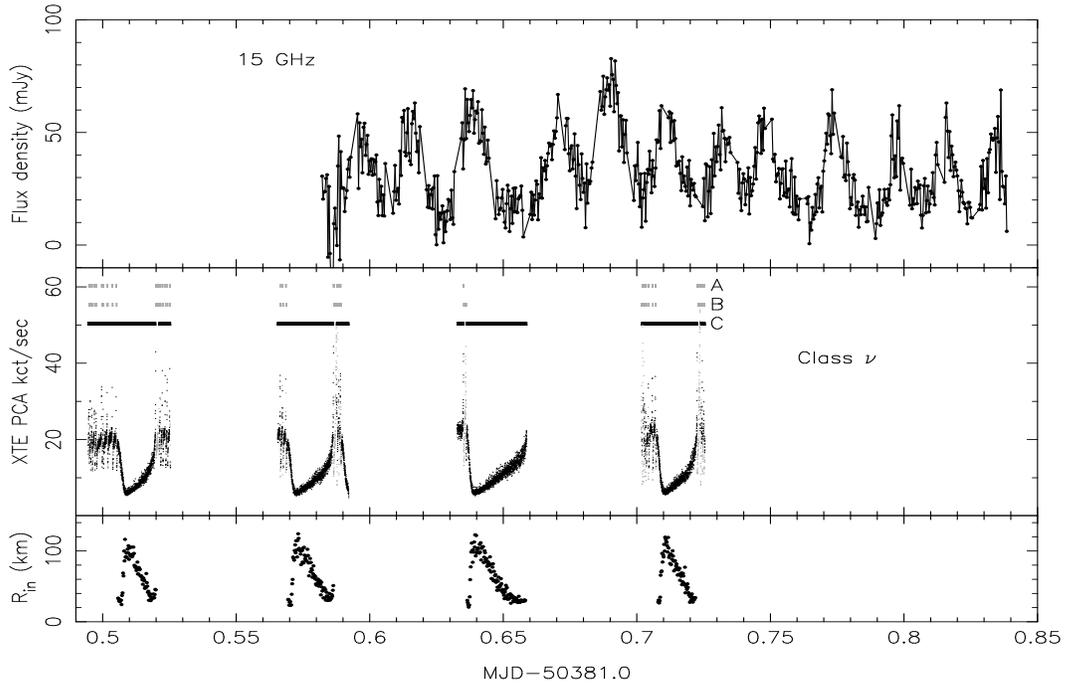}}
\caption{a -- Simultaneous radio (RT) and X-ray (\emph{RXTE} PCA) observations of GRS~1915+105 on MJD50381. The top panel shows the RT 15 GHz flux
densities plotted with 32-sec time resolution. The middle panel shows
the total X-ray count rate in the 2 -- 60 keV band (1 sec time resolution), 
broken up into States A, B and C of Belloni et al. (2000a). The darkest points are those of State C, which the analysis throughout this paper indicates is the most significant for jet formation. The class (`$\nu$') of Belloni et
al. (2000a) is also indicated. The lower panel shows the variability of
the inner accretion disc radius, R$_{in}$, derived from spectral fits to the
X-ray data. The X-ray dips, apparently caused by the `disappearance'
of the inner $\sim 100$ km of the accretion disc, appear to be
closely related to the repeated radio oscillations. Figs.~\ref{abc50681}a--\ref{abc50706}g are presented in a similar way, but notice that gray scales and both the abscissae and the ordinates ranges are not always the same.}
\label{abc50381}
\end{figure*}

Presented in Fig.~\ref{abc50381}a, this observation was first published in Pooley \& Fender (1997), and was the first direct evidence for a disc -- jet interaction on short (minutes) time-scales from an X-ray binary. The format of Fig.~\ref{abc50381}a is repeated in Fig.~\ref{abc50681}b, c, d, e, f, g and is described in detail in the caption; from top to bottem we plot:
\begin{itemize}
\item{The radio flux density at 15 GHz, as measured by the RT, in
32-second time bins.}
\item{The total \emph{RXTE} PCA count rate (with a 1 sec time resolution), additionally gray-scale coded to distinguish the three basic states (A, B \& C), with the class indicated. Emphasis is placed on Class C (when present).}
\item{Best fit values of the inner accretion disc radius (with a 16 sec time resolution), as described above.}
\end{itemize}
When comparing Figs.~\ref{abc50381}a -- g, note that the abscissae and the ordinate ranges are not always the same.

For observation MJD50381 (Fig.~\ref{abc50381}a) we observe a radio light curve which is completely dominated by oscillations with a quasi-period of $\sim 40$ min. We will refer to these oscillations as the radio oscillation events or simply as radio oscillations. The simultaneous X-ray observations reveal a light curve (classified as class $\nu$) dominated by State C dips which appear to recur with the same quasi-period as the radio oscillations (although the sparse X-ray coverage allows for the recurrence period to be twice that of the radio events, this is ruled out by observations MJD50698 and MJD51342). The (hard) State C dips imply, in the simple X-ray model, a `disappearance' of the inner $\sim 100$ km of the accretion disc, which then slowly refills. This is reflected by the changes in the inner radius, presented in the lower panel of Fig.~\ref{abc50381}a. Thus we have convincing evidence for the repeated apparent disappearance and refill of the inner accretion disc, with associated radio events.

\setcounter{figure}{2}
\begin{figure*}
\centering{\epsfig{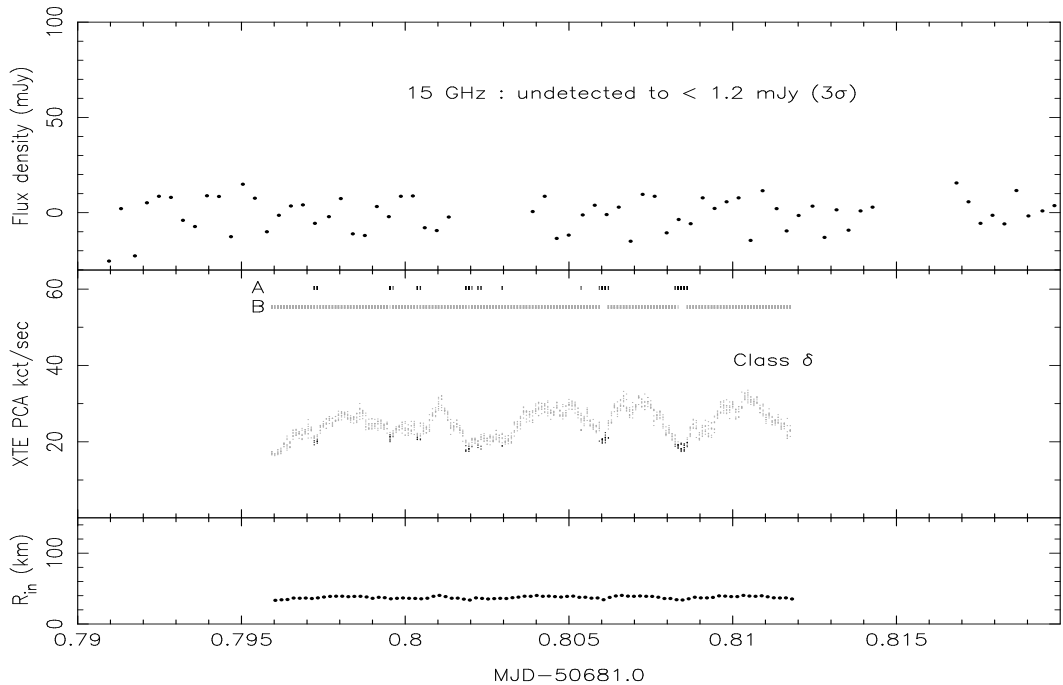}}
\caption{b -- Data presented as in Fig.~\ref{abc50381}a, with emphasis placed on state A. Note that in this case there are no State C periods, and the inner accretion disc appears to extend right down to $\sim40$ km throughout the observation; in this case changes in the flux are interpreted as being due to changes in the temperature, which in turn reflects the varying mass accretion rate through the disc. In such disc-dominated states, no radio oscillation events are observed and the radio emission is low (typically around a few mJy); in this case the radio emission is even below detectable levels.}
\label{abc50681}
\end{figure*}

\setcounter{figure}{2}
\begin{figure*}
\centering{\epsfig{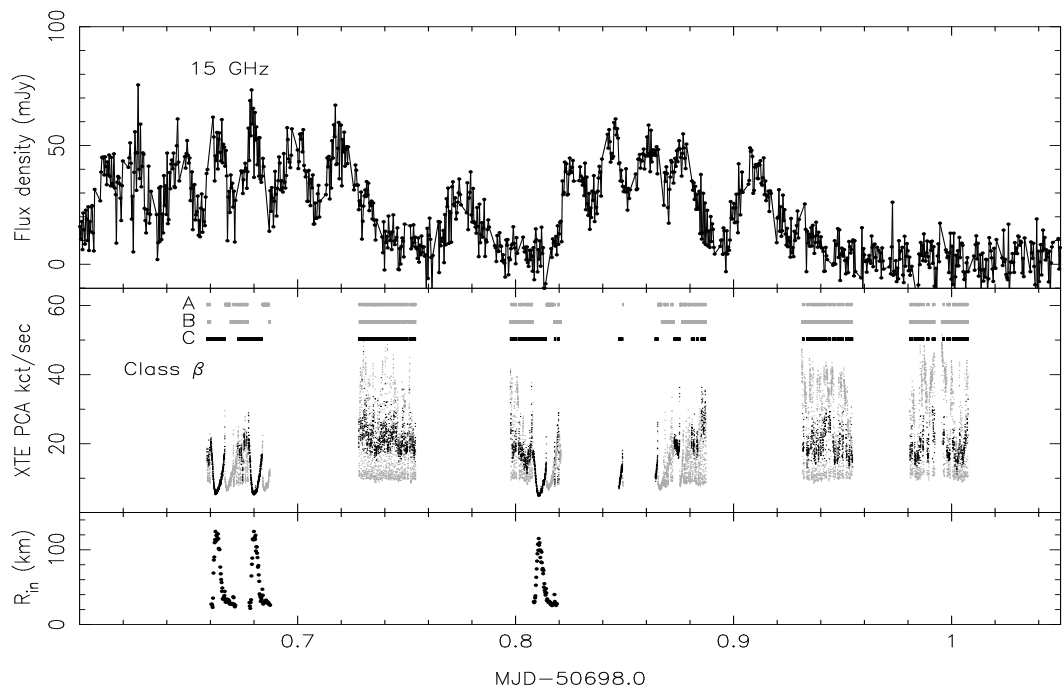}}
\caption{c -- Data presented as in Fig.~\ref{abc50381}a. In this case we see that the intermittent periods of radio oscillations appear to occur only when
there are prolonged, deep State C dips (with a $\sim$30 minutes delay, see Section~\ref{sec:onetoone}). During periods of rapid A -- B-- C flickering the radio emission drops to lower (although still
detectable) levels and shows no clear individual events. Only during
the long dips can we make spectral fits to the inner accretion disc
radius, which is behaving much as it did on MJD50381 (Fig.~\ref{abc50381}a).}
\label{abc50698}
\end{figure*}

\setcounter{figure}{2}
\begin{figure*}
\centering{\epsfig{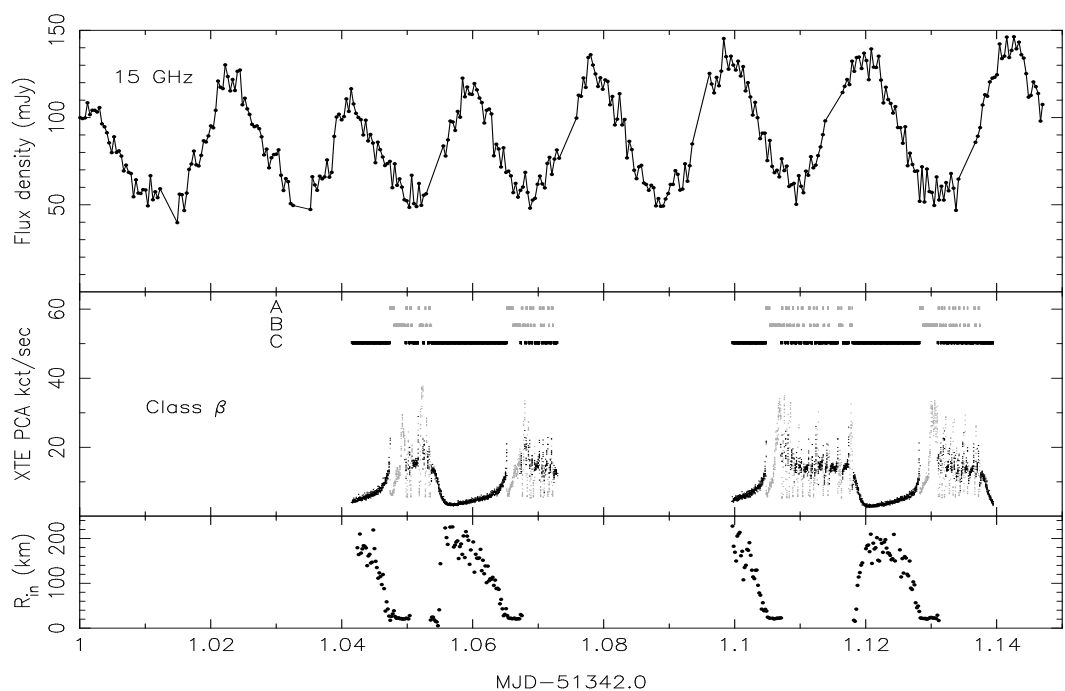}}
\caption{d -- Data presented as in Fig.~\ref{abc50381}a, however, notice the different vertical scale in the top and lower panels. In this case we again see a clear relation between long State C dips and radio oscillations. Note that in this case we can clearly see both the radio and X-ray repetition
periods, which clearly identifies one X-ray dip with one radio
oscillation. Note also that in this case the changes in the inner disc radius are more dramatic compared to MJD50381 (Fig.~\ref{abc50381}a) and MJD50698 (Fig.~\ref{abc50698}c), as R$_{in}$ reaches up to $\sim$200 km.}
\label{abc51342}
\end{figure*}

\setcounter{figure}{2}
\begin{figure*}
\centering{\epsfig{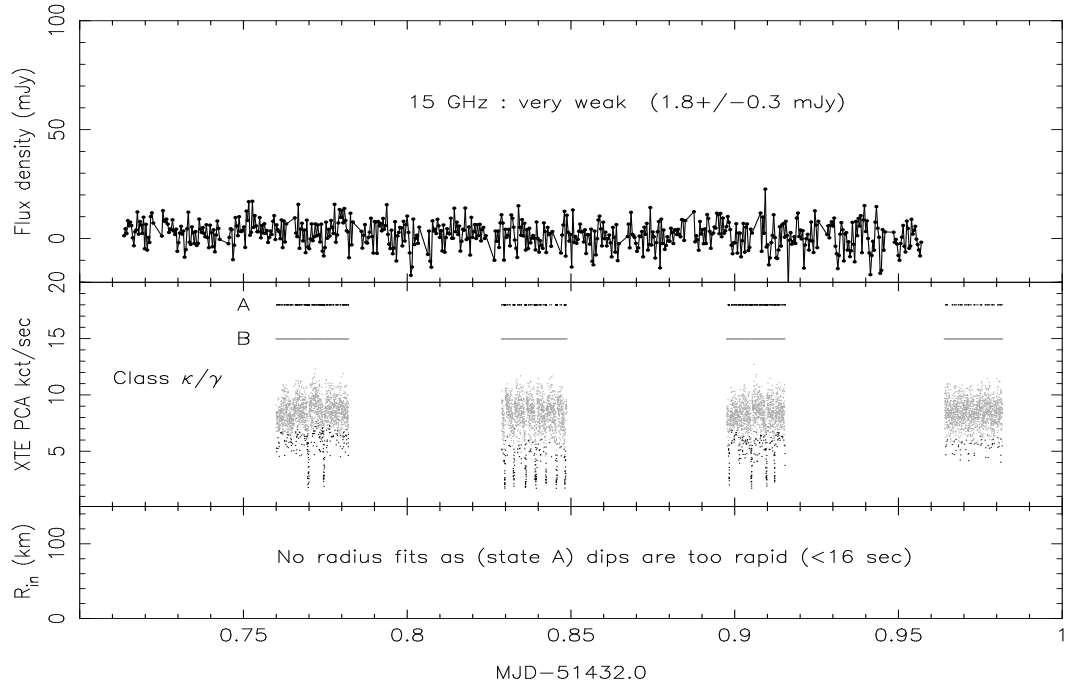}}
\caption{e -- Data presented as in Fig.~\ref{abc50381}a, however, notice that the vertical scale is different in this figure and also that the States A and B are highlighted in different colours. Although the X-ray flux shows rapidly repeating soft, State A, dips the 15 GHz radio flux is very weak. The X-ray variability is classified as belonging to class $\kappa$ except orbit numbers 2 and 3 from the third interval, these belong to class $\gamma$. Because the State A dips are shorter than 16s, no spectral fits could be performed.}
\label{abc51432}
\end{figure*}

\setcounter{figure}{2}
\begin{figure*}
\centering{\epsfig{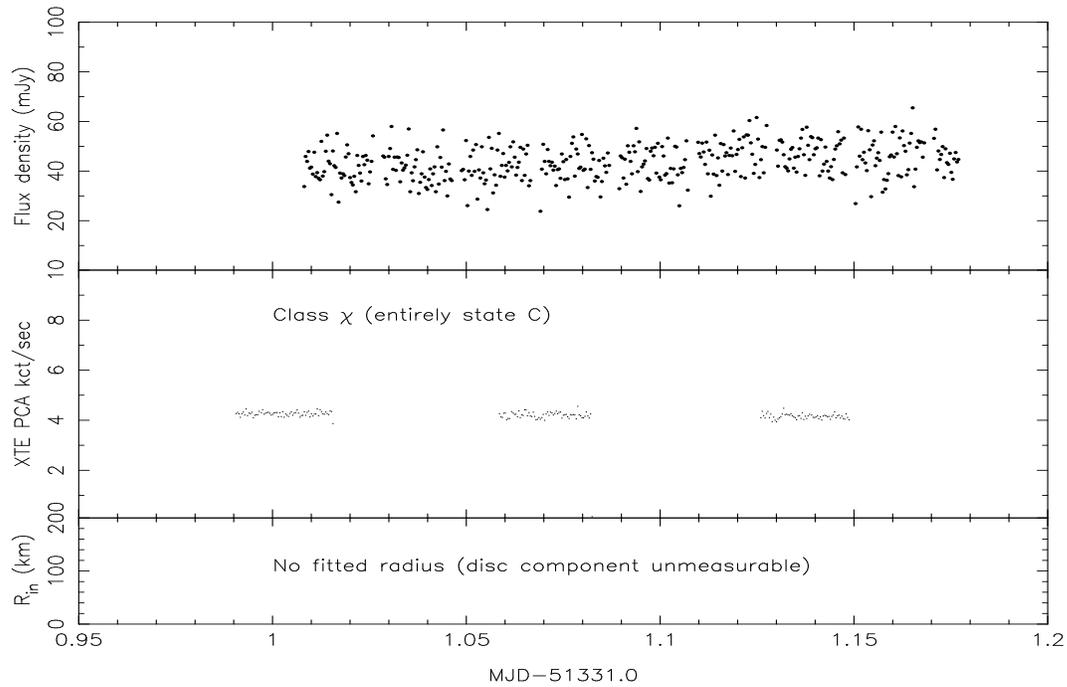}}
\caption{f -- Data presented as in Fig.~\ref{abc50381}a, however, notice the different vertical scale in the middle and lower panels. The X-ray light curve is classified as class $\chi$: the source stays in State C during the whole observation, and only little variation is found in the light curve. Not only are there no state transitions to and from State C, also the radio light curve does not show radio oscillation events. Instead the radio flux level is high ($\sim45$ mJy) and almost continuous. Because the X-ray energy spectrum is dominated by a strong hard component, the soft disc component is not well constrained which prevented the production of reliable spectral fits. In order to get an estimate of the inner radius we performed fits of all the three individual observation intervals added together, and found for the inner radius of $\sim240$ km.}
\label{abc51332}
\end{figure*}

\setcounter{figure}{2}
\begin{figure*}
\centering{\epsfig{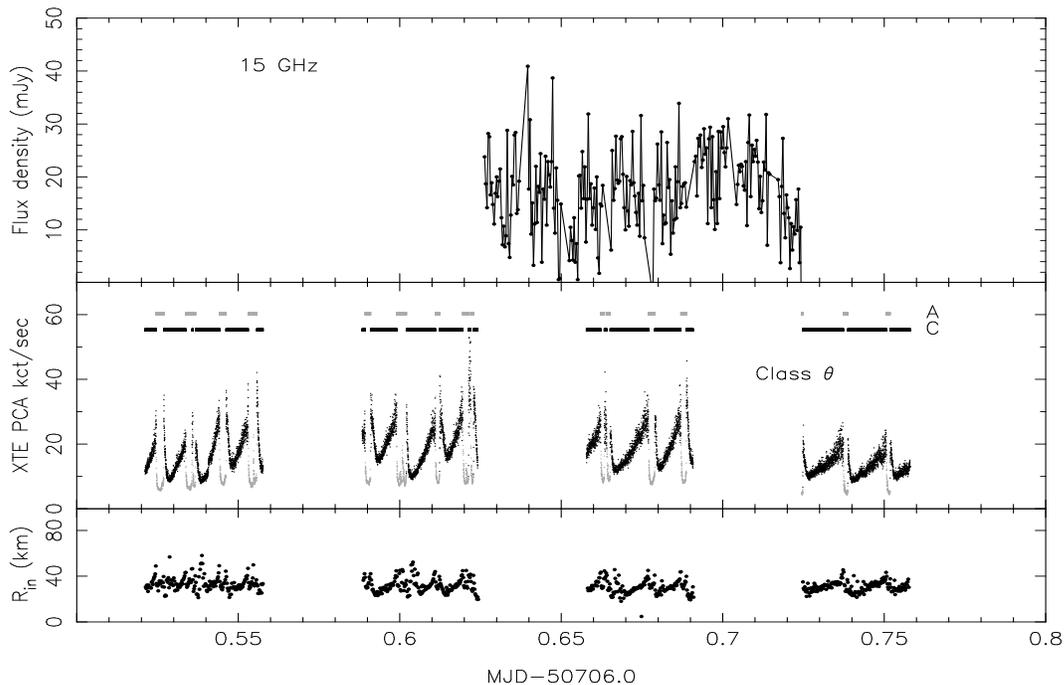}}
\caption{g -- Data presented as in Fig.~\ref{abc50381}a, however, notice the different vertical scale in the top and lower panels. This observation belongs to class $\theta$, which shows transitions between State C and A. Although the State C intervals are long enough, there are no clear radio oscillations found. Notice that the changes in the inner radius are very small compared to the classes $\beta$ and $\nu$: the inner radius of the accretion disc is almost the same for state C and A.}
\label{abc50706}
\end{figure*}

\subsection{MJD50681 (1997 August 21)}

Presented in Fig.~\ref{abc50681}b, these observations reveal a very different
relation between disc and jet in GRS~1915+105. The X-ray light curve
still displays fairly large amplitude variability (up to 30 per cent in total intensity), but it is only between States A and B, with no State C intervals. In the model of Belloni et al. (2000a) this simply corresponds to changes in the
inner temperature of an accretion disc which extends nearly down to the innermost stable orbit. This is well illustrated by the fits to the inner disc radius, which stay very close to $\sim40$ km throughout the observation. Note that the temperature changes are in turn presumed due to changes in the mass accretion rate through the disc.

Crucially, there is no radio emission detected during the whole observation: the radio flux density is very low, and below detectable levels ($\le$1.2 mJy, 3$\sigma$). This suggests that when there is a bright, spectrally soft accretion disc extending up to the innermost stable orbit and a hard coronal component is weak or absent, the radio emission is low and no radio oscillation events are produced (despite apparently significant variability in the mass accretion rate through the disc).

\subsection{MJD50698 (1997 September 7)}

The data from these long (440 min total) overlapping observations are
presented in Fig.~\ref{abc50698}c. The overall classification of the
X-ray behaviour is class $\beta$: the source clearly varies
between deep State C dips (reminiscent of MJD50381 (Fig.~\ref{abc50381}a))
and rapid A -- B -- C variations. The deep State C dips reach a
similar maximum inner disc radius of $\sim 100$ km. The pattern of
radio emission is such that during the periods with deep State C dips,
there are repeated radio oscillations with a quasi-period of around 20
min, which are delayed with respect to the X-ray emission by $\sim$30 min (see Section~\ref{sec:onetoone}). During the rapid A -- B -- C variability, however, the radio emission drops to a low level, typically $\sim$1.9~$\pm$~0.4 mJy (which is about the same level as during the State A -- B transitions seen in observation MJD51432, Fig.~\ref{abc51432}e), with no clear bright events. Note that, during the radio oscillations, the count rate of the State C dips (count rate drops to $\sim5$ kct~sec$^{-1}$) and the amplitude of the radio oscillations ($\sim 40$ mJy) are both comparable to MJD50381 (Fig.~\ref{abc50381}a). 

The preliminary conclusion that can be drawn from this simultaneous observation is that it is not just State C which is required to generate high flux radio oscillations, but that relatively long State C (longer than 100s, see Section~\ref{sec:emp}) intervals are required in order to generate observable radio oscillation events. As the long State C intervals are also both deeper and further apart than the short State C intervals, any of these three properties (or a combination of them) might be required for radio oscillation events.

\subsection{MJD51342 (1999 June 13)}

MJD51342 (Fig.~\ref{abc51342}d) displays another sequence of relatively
long State C dips followed each time by relatively long periods of A
-- B -- C variability. In this observation both the radio and the X-ray repetition periods are clearly visible from the light curves. For the X-ray dip recurrence time we find $\sim 30$ min. The radio emission shows a very regular sequence of radio oscillations with the same recurrence quasi-period, unambiguously associating one radio event with each dip. It is, however, still not clear to \emph{which} dip each of the radio events is related. The X-ray spectral fits indicate that the changes in the disc inner radius are more dramatic compared to MJD50381 (Fig.~\ref{abc50381}a) and MJD50698 (Fig.~\ref{abc50698}c), increasing up to $\sim 200$ km. 

\subsection{MJD51432 (1999 September 11)}
\label{sec:noA}

The X-ray behaviour on MJD51432 is classified as class $\kappa$/$\gamma$ (see Fig.~\ref{abc51432}e). These classes are characterized by \emph{soft}, State A dips in the light curve which are much deeper than the State A dips in observation MJD50681 (about a factor 10 in count rate lower, see Fig.~\ref{abc50681}b). Despite the relatively large X-ray variability, corresponding to the State A -- B transitions, the radio emission is low ($\sim$1.8 mJy). These observations, together with the ones shown in Fig.~\ref{abc50681}b (MJD50681), clearly indicate that (soft) State A and B intervals can \emph{not} be related to radio oscillation events, but instead are characterized by a low-level radio emission.

\subsection{MJD51332 (1999 June 3)}

The previous examples have shown either the case when there is a low level radio emission and no radio oscillation events, or when there is a high radio flux and radio oscillation events. In Fig.~\ref{abc51332}f we show an example of an observation which has a relatively high level of radio flux, but no radio oscillations. The X-ray light curve is classified as class $\chi$ and shows very little variations and/or state transitions. In fact, the source stays in State C for periods of days to months. From these observations we conclude that an uninterrupted State C interval can produce a high level of radio emission, but \emph{no} radio oscillation events. However, note that there are a considerable number of uninterrupted State C intervals with low radio flux ($\sim5$ -- $10$ mJy), also without radio oscillation events (see Table~\ref{t1}). 

As explained in Section~\ref{sec:xrayspectra} we can only get a rough estimate of the inner radius of the accretion disc from the spectral fits due to the fact that the soft component is not well constrained. However, from spectral fits of the whole observation (all intervals added) we find for the inner radius of the accretion disc $\ga240$ km. This suggest that the inner radius is much larger during the class $\chi$ observations than during the long (interrupted) State C intervals. 

\begin{table*}
\caption{The Linear correlation coefficient and the Spearman Rank correlation coefficient for the correlations between the RT radio flux and the \emph{RXTE} ASM flux and HR$_{2}$ colour, as shown in Fig.~\ref{statec}. A positive value of the correlation coefficient indicates a \emph{correlation}, and a negative value an \emph{anticorrelation}. For each of the correlations the confidence level (in percent) is given.\label{t3}}
\begin{tabular}{c|cc|cc|cc|}
\hline
Radio vs. X-ray	& \multicolumn{2}{|c}{$\chi_1$} & \multicolumn{2}{|c}{$\chi_2$} & \multicolumn{2}{|c}{$\chi_3$} \\
	& Coeff.  & Conf.  & Coeff. & Conf. & Coeff. & Conf.\\
\hline
Lin. corr. coeff. &  $-0.63$ & $>99$\% &$+0.52$ & $>99$\% & $-0.87$ & $>99$\%  \\
Rank corr. coeff. &  $-.044$ & $\sim~98$\% & $+0.63$ & $>99$\% & $-0.82$ & $>99$\% \\
\hline
\end{tabular}
\begin{tabular}{c|cc|cc|cc|}
Radio vs. HR$_{2}$	& \multicolumn{2}{|c}{$\chi_1$} & \multicolumn{2}{|c}{$\chi_2$} & \multicolumn{2}{|c}{$\chi_3$} \\
	& Coeff.  & Conf.  & Coeff. & Conf. & Coeff. & Conf.\\
\hline
Lin. corr. coeff. &  $+0.47$ & $\sim~99$\% &$-0.14$ & $\sim~68$\% & $+0.88$ & $>99$\%  \\
Rank corr. coeff. &  $+0.38$ & $\sim~95$\% &$-0.15$ & $\sim~79$\% & $+0.84$ & $>99$\% \\
\hline
\end{tabular}
\end{table*}

\subsection{MJD50706 (1997 September 15)}
\label{theta}

The observation presented in Fig.~\ref{abc50706}g is classified as class $\theta$ and shows transitions between State A and C. When we compare this observation with the ones presented in Fig.~\ref{abc50381}a, c and d we immediately notice two major differences. First of all, although the State C intervals in class $\theta$ are long enough ($\ga100$s, they are comparable in length to the State Cs in class $\beta$ and $\nu$) and seem to be related to a relatively high level of radio flux (see Table~\ref{allcl}), only \emph{strong variability} (with no periodicity) is found in the radio flux and clear oscillation events such as found in class $\beta$ and $\nu$ are absent. Secondly, the radius fits do not show the same pattern (fast increase in the inner radius, followed by a slower decrease) during the State C intervals we found in class $\beta$ and $\nu$. In fact we find that the inner radius shows no large variations and the value we find in State C is comparable to that in State A. 

From Table~\ref{t1} and \ref{allcl} it is clear that class $\theta$ is the only class which shows long State C intervals (like the ones from class $\beta$ and $\nu$ which are related to the radio oscillation events), but no State B intervals. Another difference between class $\theta$ and classes $\beta$ and $\nu$ is that the long State C dips in class $\theta$ are near-contiguous, with only very little time ($\la3$ min) between them, whereas in classes $\beta$ and $\nu$ this is much larger ($\sim20$ min). So, another possibility is that the additional requirement on State C dips to produce radio oscillation events is that they are ``well-separated''.

\begin{figure*}
\centering{\epsfig{figure=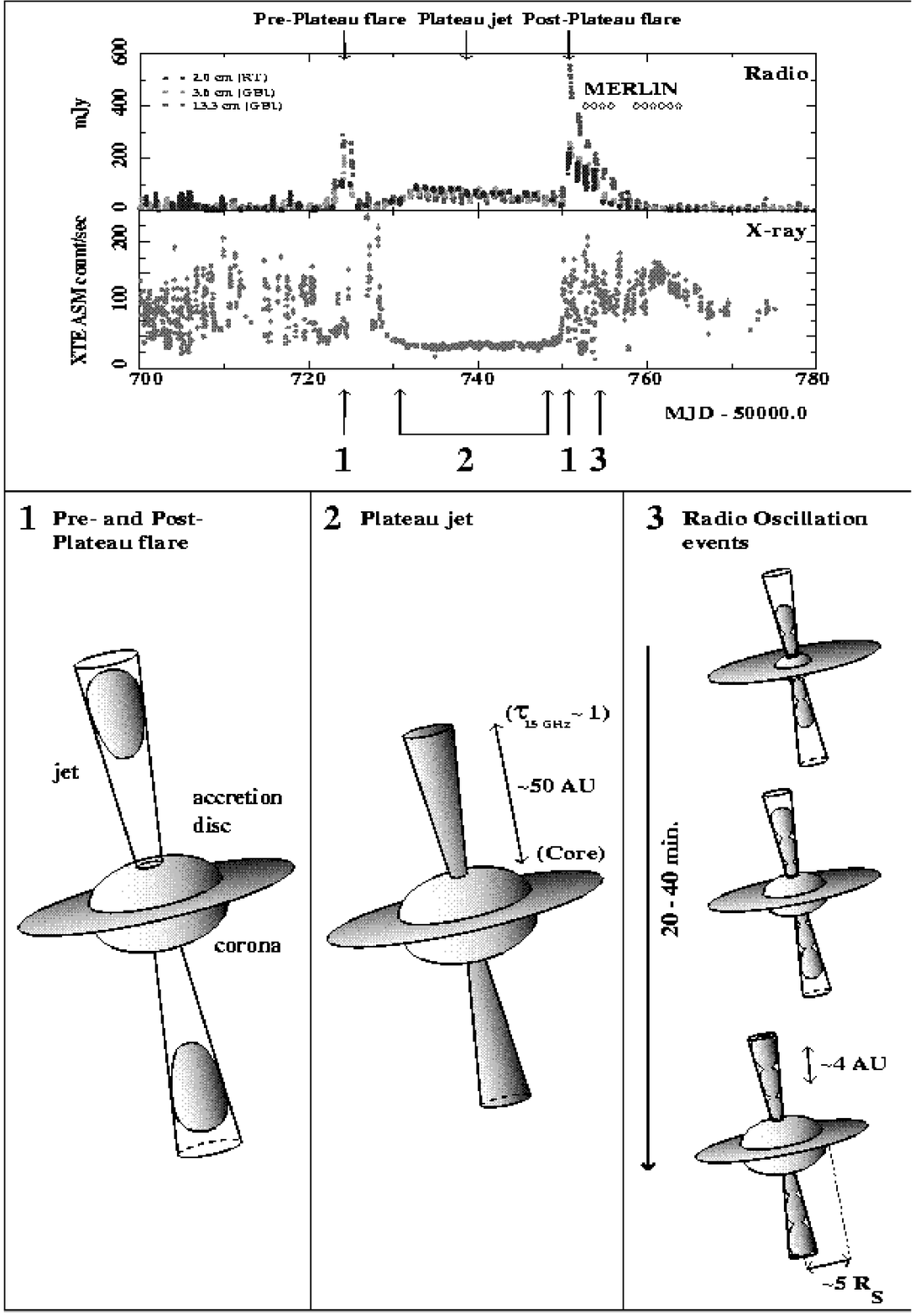, width=13cm,height=18cm}}
\caption[]
{A schematic view of the radio oscillation events, the pre- and
post-plateau flares and the plateau jets in GRS~1915+105. In the top
panel we show the radio (RT and GBI) and the X-ray (\emph{RXTE} ASM)
light curves from MJD50700 to MJD50780. We have indicated the
positions of the flares, the jet and also of the start of the radio
oscillation events (which start just after the beginning of the
post-plateau flare). Also indicated are the times the source was
observed with MERLIN (see also Fender et al. 1999b). In the lower
panel we show schematic views of the ejection events we encountered
(the numbers correspond to the positions in the light curve indicated
in the top panel): 1) the pre- and post-plateau flares; 2) the plateau
jet, with quasi-continuous ejections and no single blobs observed; 3)
the radio oscillation events, a sequence (in the direction of the
arrow) is shown during which a new blob is ejected. The figures are
not drawn to scale and estimates for the sizes of the jets and the
corona are indicated. R$_{s}$ is the Schwarzschild radius, and we
assumed a black hole mass of 10~M$_{\sun}$. For the pre- and
post-plateau flares the blobs are observed up to distances of
$\sim3000$ AU from the core (Fender et al. 1999b, Dhawan et
al. 2000). In the case of the plateau jet the distance to the
optically thin point ($\tau$=1) at 15 GHz is given (see also Dhawan et
al. 2000), and for the radio oscillation events a typical distance
(upper limit) between two blobs is estimated from the period of the
oscillations ($\sim30$ min) and assuming the blobs are moving with the
speed of light. Notice that during the radio oscillation event
sequence each new ejection starts before the previous one has
completely faded away, as can be found from the light curves
(Figs.~\ref{abc50381}a, c, and d).}
\label{fig:jets}
\end{figure*}

\section{Class $\chi$ observations}
\label{sec:platstates}
\begin{figure*}
\centering{\epsfig{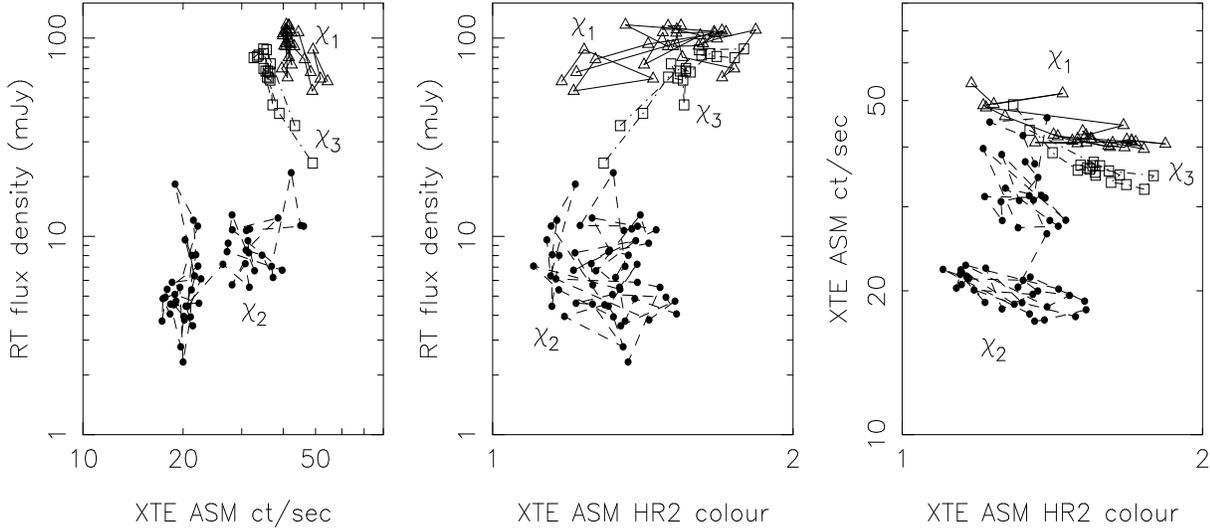}}
\caption[]{From left to right: the 15 GHz radio flux density as a function of \emph{RXTE} ASM 2 -- 12 keV flux, the 15 GHz radio flux density as a function of the hardness ratio (HR$_{2}$=$\frac{3 - 5 keV}{5 - 12 keV}$), and the \emph{RXTE} ASM 2 -- 12 keV flux as a function of the hardness ratio, using daily-averaged data for the long State C intervals $\chi_1$, $\chi_2$ and $\chi_3$ (Belloni et al. 2000a). Note that for $\chi_1$ and $\chi_3$ there is \emph{anticorrelation} between radio flux density and total X-ray flux, and a \emph{correlation} with hardness. For $\chi_2$ the situation is reversed: the radio emission is \emph{correlated} with X-ray flux and \emph{anticorrelated} with hardness. For all the three $\chi$ intervals there seems to be an anticorrelation between the X-ray flux and the hardness ratio: when the source becomes brighter, it gets softer.}
\label{statec}
\end{figure*}

Besides the long State C intervals which are related to the radio oscillation events, there are also times when the source occupies State C for periods of days to months. Observations during these periods are marked ``entirely C'' in Table~\ref{t1} and are characterized by steady radio and X-ray emission. An example of such an observation is shown in Fig.~\ref{abc51332}f. Belloni et al. (2000a) classified all these observations, available to them at that time, as class $\chi$ and subdivided them, as a time sequence, in the subclasses $\chi$$_{1}$, $\chi$$_{2}$, $\chi$$_{3}$. Below we will discuss the three subclasses in more detail. Also a fourth subclass, $\chi$$_{4}$, was introduced which comprises all the observations that did not fall in one of the three other classes. As this subclass does not consists of a time--sequence of state C observations, it will not be discussed as a separate class below.

\subsection{The radio behaviour during the $\chi$ classes}
\label{sec:radiobehaviour}
The intervals $\chi_{1}$ and $\chi_{3}$ are referred to as ``plateau'' states (or as radio-loud, radio-plateau (Muno et al. 2001), or \emph{type~II} states (Trudolyubov 2001)) and a handful of these states have been observed since 1996 (Pooley \& Fender 1997, Fender et al. 1999b, Fender 2001). The pattern of these ``plateaux'' seems fairly stable (Pooley \& Fender 1997; Hannikainen, Hunstead, Campbell-Wilson 1998; Fender et al. 1999b, Fig.~\ref{fig:jets}). They are generally preceded by (an) optically thin radio event(s) (``pre-plateau flare''), settle into the quasi-stable state for a few weeks (``plateau jet''), and then are followed by further optically thin radio event(s) (`post-plateau flare'; see also Fig.~\ref{fig:jets}). During the quasi-stable state the radio emission seems to saturate at $\sim$100 mJy (at 15 GHz; Fig.~\ref{statec}), while the radio flux during pre-plateau and the post-plateau flares generally is a factor 3 to 6 higher (see top panel Fig.~\ref{fig:jets} and Mirabel \& Rodr\'\i guez 1995, Hannikainen et al. 1998 and Fender et al. 1999b). It is only for $\chi_3$ that the quasi-continuous ejections and the discrete ejections corresponding to the quasi-continuous radio jet and the radio flares respectively are directly observed (Dhawan et al. 2000). 

During class $\chi_2$ the radio flux is typically a factor $\sim10$ lower compared to $\chi_{1}$ and $\chi_{3}$ (see Fig.~\ref{statec}), and no major ejections of relativistic plasma have been reported. Hence the $\chi_2$ interval is often referred to as radio-quiet (Muno et al. 2001; or as \emph{type~I} (Trudolyubov 2001)).

Simultaneous infrared coverage in 1996 (Bandyopadhyay et al. 1998) indicates that the flat ($\alpha$$\sim0$, at GHz frequencies) radio spectrum during the plateau states may extend to near-infrared wavelengths (Fender 2001). In Fender et al. (2000) the flat-spectrum emission of GRS~1915+105 during the plateau states is compared to that of Cyg~X-1 and Cyg~X-3, and it is argued that such spectra arise in a conical, quasi-continuous, partially self-absorbed outflow such as originally envisaged (in an idealized case) by Blandford \& K\"{o}nigl (1979) and developed by others, including application to X-ray binaries (e.g. Reynolds 1982; Hjellming \& Johnston 1988; Falcke \& Biermann 1996, 1999). In this case, the plateau states correspond to the formation of a luminous quasi-steady outflow from GRS~1915+105, a scaled-up version of the jet from Cyg~X-1 in the Low/Hard X-ray state (Stirling et al. 2001). 

\begin{figure}
\centerline{\psfig{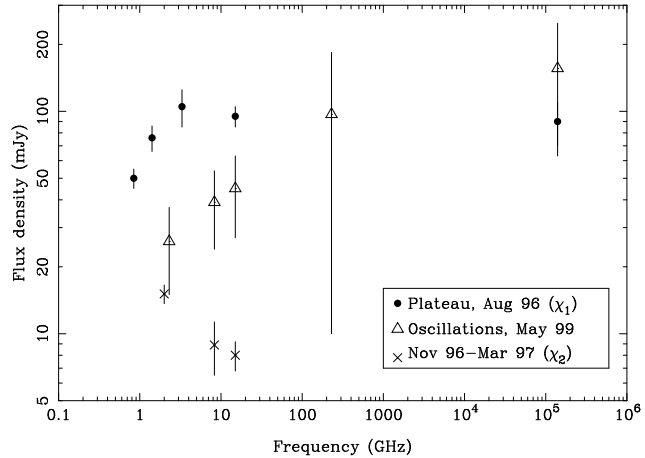}{\hfil}}
\caption[]{The radio -- infrared spectra for the plateau jet, the $\chi_2$ interval and the radio oscillation events. The $\chi_2$ spectrum is the first reported radio spectrum for this interval. The spectrum is flat, with a spectral index of $\alpha$=$-0.2\pm0.1$. See Section~\ref{sec:radiobehaviour} for more details. The spectrum of the plateau interval is taken from Fender (2001), and the spectrum of the radio oscillation events is taken from Fender \& Pooley (2000).}
\label{fig:platosc}
\end{figure}

For class $\chi_{2}$ we combined the GBI and Ryle telescope averaged data points to obtain the first reported radio spectrum during $\chi$$_{2}$. The flux at 2.3 GHz showed large variations as a function of the Hour Angle (HA), while the 8.3 GHz flux stayed almost constant. The larger field of view at 2.3 GHz may result in a significant contribution to confusion from nearby sources, and this may result in the fluctuations noted. We therefore use the data at 8.3 and 15 GHz. The average radio spectrum is plotted in Fig.~\ref{fig:platosc}, and for the spectral index (between $8.3$ -- $15$ GHz) we find: $\alpha$=$-0.2\pm0.1$. This means that although during $\chi_{2}$ no radio jets or radio oscillations were observed and the radio flux was significantly lower compared to classes $\chi_1$ and $\chi_3$, the (high frequency) radio spectrum is also found to be approximately flat.

\subsection{X-ray and radio correlations}
\label{sec:x-ray:radio}

In Fig.~\ref{statec} we compare daily-averaged radio and X-ray data for the prolonged State C periods $\chi_1$, $\chi_2$ \& $\chi_3$. In Table~\ref{t3} we show both the Linear and the Spearman Rank correlation coefficients (Press et al. 1992) and their confidence level for all the relations shown in Fig.~\ref{statec}. From this we can conclude that for the plateau states $\chi_1$ and $\chi_3$:
\begin{itemize}
\item{the radio flux density is \emph{anticorrelated} with the soft X-ray flux;}
\item{the radio flux density is \emph{correlated} with the ASM HR$_{2}$ colour (for definition see caption Fig.~\ref{statec}).}
\end{itemize}
And for the state $\chi_2$ we find:
\begin{itemize}
\item{the radio flux density is \emph{correlated} with the soft X-ray flux;}
\item{the radio flux density may be \emph{anticorrelated} with the ASM HR$_{2}$ colour.}
\end{itemize}
So while the plateau states $\chi_1$ and $\chi_3$ show very similar
behaviour with respect to the radio and X-ray correlation, the $\chi_2$ interval shows a different behaviour. Note that this division is also reflected in the X-ray behaviour: $\chi_2$ has a lower count rate compared to $\chi_1$ and $\chi_3$ (Fig.~\ref{statec}). Also Belloni et al. (2000a) found that $\chi_2$ occupies a lower branch in the CD compared to $\chi_1$ and $\chi_3$, a phenomenon they explained as related to a difference in N$_{H}$ and/or a difference in the high energy cutoff.

\section{Empirical features of the disc -- jet connection}
\label{sec:emp}

\subsection{State C intervals}
\label{sec:statecinter}
In order to check that indeed we need a long, single State C interval to produce a high level of radio emission, and not just `a lot of State C' we calculated the time the source was in State C during the intervals when no long State Cs where present for observation MJD50698 (Fig.~\ref{abc50698}c). We compared this with the time the source shows long State C intervals (for the same observation). We find that for the intervals around MJD50698.67 and MJD50698.81, which are associated\footnote{Here we used the fact that the radio emission is delayed with respect to the X-ray emission, see Section~\ref{sec:onetoone}} with a high radio flux (and radio oscillation events), the source shows long State Cs for 45 percent and 27 percent of the total time, respectively. On the other hand, the intervals around MJD50698.74, MJD50698.94, MJD50698.98 and MJD50699.0, which are related to a low-level radio emission, show short State C for 41 percent, 30 percent, 29 percent and 31 percent of the total time, respectively. This clearly shows that the presence alone of State C is not enough: there are at least four observation intervals which lack the long State Cs, the high radio emission, and the radio oscillation events, but spend more or a comparable fraction of their time in State C compared to the observation intervals which do show long State Cs. It also shows that the long State C intervals have a disproportional higher level of radio flux compared to the short State C intervals.

Is it only the length of the State C intervals that determines if the radio oscillation events are produced? From Section~\ref{sec:obs} and Table~\ref{t1} it is clear that classes $\theta$ and $\alpha$ do show long State C intervals and a relatively high level of radio flux, but no clear radio oscillation events. Unfortunately, due to the poor quality of the X-ray~:~radio overlap in the class $\alpha$ observations, we were not able to say anything about the behaviour of the source in this class. For class $\theta$ we do have good overlaps, and Fig.~\ref{abc50706}g shows an example. From Section~\ref{theta} and the above, it seems that for the high flux radio oscillation events to be observed, two conditions have to be met:
\begin{itemize}
\item{the State C intervals have to be longer than $\sim100$s;}
\item{the State C intervals have to show a repeating pattern in which they are either separated by State B intervals, or ``well-separated'' in time, or both;}
\end{itemize}
As can be seen from the Figs~\ref{abc50381}a, c and d the long State C intervals are also the ones which have the lowest X-ray count rate. Therefore, the first statement that \emph{long State C intervals} are related to the radio oscillation events also implies that \emph{low-count rate State C} intervals are related to the radio oscillation events. It would take observations of short State C intervals of low count rate to decide between these two possibilities, but such events have never been observed yet. Regarding the second statement, it is important to note that it is not clear at this point why we do see radio oscillation events in classes $\beta$ and $\nu$ and not in class $\theta$: it can be due to the presence (or absence) of the soft State B intervals, or to the larger (or smaller) separation between the State C intervals, or both. Although we feel that the presence of State B intervals can be very important (see Section~\ref{sec:longshort}), the most simple solution is that the State C intervals need to be well-separated for the individual radio events to be observed. 

It is interesting to note the difference between State Cs of class $\theta$ and the State Cs of the other variability classes. Besides the differences already mentioned in Section~\ref{theta}, we also find class $\theta$ to be different in several other ways. From comparing the lightcurve in Fig.~\ref{abc50706}g with the ones from class $\beta$ and $\nu$ (Fig.~\ref{abc50381}a, c and d) it is clear that not only do the State C intervals of class $\theta$ have the highest count rates compared to the State Cs in other classes, but the count rates are also higher than State A intervals. Regarding the spectral parameters, the class $\theta$ State Cs are also different. From the spectral fits we found an inner radius of the accretion disc which is about 3 times smaller compared to the State Cs in classes $\beta$ and $\nu$ (see Figs.~\ref{abc50381}a, c, d and g), while the inner temperature is comparable to what is found during State A and B ($\sim1.5$ keV). Together with the fact that the power law component is very steep ($\Gamma\sim3$), we would conclude that these State Cs are spectrally softer. However, from the best fit parameters we find that on average the luminosity\footnote{Luminosities are calculated assuming a distance of 11 kpc.} of the power law component ($\sim2.2$$\times$10$^{38}$erg~s$^{-1}$) is larger than the luminosity of the disc blackbody component ($\sim1.9$$\times$10$^{38}$erg~s$^{-1}$). For the State C intervals in classes $\beta$ and $\nu$ on the other hand, the power law luminosiy is smaller ($\sim0.7$$\times$10$^{38}$erg~s$^{-1}$) compared to the disc blackbody luminosity ($\sim2.5$$\times$10$^{38}$erg~s$^{-1}$). Although it is obvious that the State C intervals are clearly different in class $\theta$ compared to the other classes and we cannot rule out the possibility that these differences are responsible for the ``anomalous'' radio behaviour, more analysis is necessary to say something more firm.

\subsection{State A and B intervals}
One can ask the question whether it is only long State C intervals which produce radio oscillation events, or if perhaps long state A and B intervals do the same; i.e. is it perhaps only the length of the interval which determines the production of the radio oscillation events? The strongest evidence against long State B intervals being responsible for the production of radio oscillation events is shown in Fig.~\ref{abc50681}b. In this case we clearly have long (up to $\sim30$ min) state B intervals, but the radio emission is below detectable levels. The only cases where we have long state A intervals are during the observations presented in Figs.~\ref{abc50698}c, d and g. These observations are classified as class $\beta$ and $\theta$ and in fact it is only for observations in these classes and class $\phi$ (which consists of State A only, but for which we do not have any overlapping radio observations) that the source shows long State A intervals. In classes $\beta$ and $\theta$ each long State C interval is followed by a (slightly shorter, $\sim80$ -- $100$s) State A interval. Hence, one might relate these long State A intervals to the radio oscillation events. However, when we compare the observations presented in Figs.~\ref{abc50698}c and d with the class $\nu$ observation shown in Fig.~\ref{abc50381}a, which does not have long State A intervals, we see the same pattern of radio oscillation events (although the pattern in Fig.~\ref{abc50706}g is not so clear, we do believe the State C intervals are responsible for the variations in the radio flux; see above). We therefore conclude that only ($\geq$100s) long, well-separated \emph{State C} intervals produce radio oscillation events.

What is the difference in the radio behaviour when the X-ray light curve consists of short State C dips and when it consists of short State A dips? From Fig.~\ref{abc50381} and Table~\ref{t1} it is clear that in both cases the radio emission is low. We find that the average flux level corresponding to intervals showing short State A dips reaches $2.0\pm0.10$ mJy while for the intervals showing short State C intervals this is $3.3\pm0.10$ mJy (numbers calculated from Table~\ref{t1}). From this we conclude that, although they are clearly not related to the radio oscillation events, short State A intervals and short State C intervals show a very similar radio behaviour. In fact \emph{State A, B and short State C intervals are all related to low-level radio emission.} For some observations we see a slow decay in the low-level radio flux and therefore, at this point we cannot rule out the possibility that the radio emission is the decay from a large radio event prior to the observation in question, as these radio events can be very short ($\la24$h).

\subsection{The one -- to -- one relation; estimating the time delay}
\label{sec:onetoone}
\begin{figure}
\centerline{\epsfig{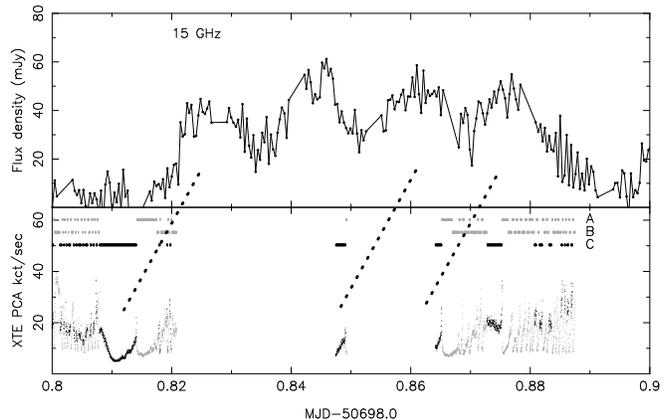}}
\caption{Close-up of radio and X-ray light curves on MJD50698,
indicating our best guess at the one-to-one correspondence between
State C dips and radio oscillation events (see also Fig.~\ref{abc50698}c).}
\label{corresp50698}
\end{figure}

It is clear from the figures that the X-ray and radio events do not occur simultaneously. In Fig.~\ref{abc50698}c the radio oscillations switch `on' and `off', allowing us to relate each State C interval with one radio oscillation event. This is indicated in Fig.~\ref{corresp50698} where we show a close-up of the X-ray and the radio light curves of MJD50698. Despite the sparse coverage in the X-ray due to earth occultations, in view of the known characteristic repeatable X-ray behaviour we believe there is convincing evidence for a one-to-one relation (as indicated in Fig.~\ref{corresp50698}) between the dips and the radio oscillation events. Moreover, we can also associate each X-ray dip with one of the radio oscillation events occurring \emph{after} this X-ray dip. This allows us to relate each of the State C intervals in the X-ray light curves to one radio oscillation event in the other observations presented here. 
A strong indication that the one-to-one relation is indeed a intrinsic feature of the source and not just a coincidence, is the fact that the periods found in the radio oscillation events are the same as the periods of the long hard State C dips in the X-ray light curve: we find periods around 30 minutes in good agreement with the range of $20$ -- $40$ min found by Pooley \& Fender (1997). Furthermore, the one-to-one relation allows us to ``fill in the gaps'' in the X-ray data which are due to earth occultations. If we assume the source shows a continuous behaviour in each class and we can safely fill in the missing gaps, this does not lead to any inconsistencies. An exception to this seems to be class $\theta$, however, we do feel that also in this case each State C interval is responsible for an ejection event and the subsequent radio emission, although we can not distinguish between these individual events (as explained in Section~\ref{sec:statecinter}).

The one-to-one relation allows us to estimate the time delay between the radio and the X-ray emission. From Fig.~\ref{corresp50698} we find a delay of $\sim$30 min; similar delays (14 -- 30 min) were also found in the other observations presented here. These delays are in agreement with theory: the self absorption in the optically thick region of the jet causes the radio emission to be delayed compared to higher frequencies, e.g. IR. Note that we measure this time delay from the end of the State C to the peak of the first subsequent radio event. These reference points were chosen because these are easily determined from the figures. This way we can only get a \emph{lower limit} of the time delay, as the real starting point of the ejection (which does not have to be the end of State C) can not be determined. Note that measuring the time delay from the beginning of State C to the peak of the subsequent radio event would make the delay longer by $\sim10$--$15$ min, which is the average length of the State C intervals.

\section{Oscillation events -- delays and profiles} 
\label{sec:delays}

By comparing simultaneous radio and IR observations Fender \& Pooley
(1998) found that most likely the IR precedes the radio by $\sim$30
min, which is about the same time delay as found for the X-ray and
the radio emission (Section~\ref{sec:onetoone}). This infers that the delay between the X-ray and the IR is small, a result already found by Mirabel et
al. (1998). Mirabel et al. (1998) fit the frequency-dependent time
delays from the near-infrared to radio wavelengths with a simple van
der Laan (1966) model. That model predicts the time delay to peak
emission at frequency $\nu$ with a function $t_{\nu} \propto
\nu^{-2/3}$  (depending on the electron distribution index) where $t_{\nu}$ is the time delay. In an idealized conical outflow, the characteristic size scale varies with frequency as $\nu^{-1}$ (e.g. Blandford \& K\"{o}nigl 1979). If the
flow is already accelerated to its terminal velocity by the time we
observe the oscillation events then the time delay $t_{\nu} \propto
\nu^{-1}$. For nearly all conceivable models in which optical depth
effects cause the frequency-dependent delays in the oscillations, the
infrared emission is likely to arise within seconds of whatever is the
disc trigger event (due to lower optical depth at higher
frequencies). Therefore, directly associating jet formation with some
particular part of the X-ray dip may be more profitably pursued at
infrared wavelengths. However, the following should be borne in mind
with respect to the emission profiles observed at different
wavelengths:

\begin{itemize}
\item{The similarity of the oscillation profiles at radio and infrared
wavelengths, and the overall flat radio--infrared spectrum, both
indicate that the self-similar phase of synchrotron source evolution
has already been reached by the time the infrared emission is
observed.}
\item{This implies that optical depth effects do not play a role in
determining the observed profiles of the oscillation events, and that
all IR--mm--radio events are observed from material already moving
at, or near to, the terminal velocity of the outflow.}
\end{itemize}

If the rise phase of the oscillations is not an optical depth
effect, it probably either reflects:
\begin{enumerate}
\item{The period of injection of relativistic particles into
the outflow from the disc/corona}
\item{The `width' of a shock which is passing down the jet
accelerating particles {\em in-situ} (e.g. Kaiser, Sunyaev, Spruit 2000).}
\end{enumerate}
In case (i) we may well be able to relate the rise timescale to
some timescale in the X-ray light curve; in case (ii) this is not so
clear, since we do not understand how the shock would be generated. However, in both of the above cases (i and ii), the lack of optical depth effects in the rise profile indicates that the timescale for particle injection/acceleration is long compared to the time required for the flow to pass through its $\tau=1$ surface (``photosphere'').

Moreover, based on the similarities in the oscillation profiles at different wavelengths, the decay phase probably reflects the intrinsic adiabatic decay of the expanding cloud which is also ``smeared out'' due to the relativistic Doppler effects. As the decay phase is not related to the X-ray behaviour a comparison of the IR-mm-radio decay phase with the X-ray light curve is not relevant (regardless of the Doppler effects). 

All of this leads us to conclude that:
\begin{itemize}
\item{we may well be able to usefully compare the rise time of the radio oscillation events with some timescales observed in the X-ray light curve;}
\item{both the rise phase (particle injection and/or acceleration by passing shock) and decay phase (adiabatic expansion losses) are frequency-independent effects, which nicely explains why the profiles are identical at all
frequencies.}
\end{itemize}

We note three caveats to the above line of reasoning -- firstly, at lower radio frequencies the oscillations do appear to become a bit less sharp, perhaps indicating the optical depth effects are beginning to play a role (see e.g. Mirabel et al. 1998); secondly, in at least one model the hard X-rays are also beamed (Markoff, Falcke \& Fender 2001), although not by a very large factor; thirdly, besides the adiabatic expansion the decay phase may additionally be affected by the decay of the injection of material into the jet (which is not affected by the Doppler effects). The above line of reasoning may well help to constrain the physical dimensions and optical depth structure of the jet in GSR 1915+105 and other systems. 

It is worth making a further clarification here. When we state
`ejection' or `injection' we are talking about a population of {\em
relativistic} particles which we observe via their synchrotron
emission. We cannot rule out possibilities such as a continuous
outflow of particles, which in soft states (A and B) are cooled before we
can observe the synchrotron emission. This cooling may occur via the inverse Compton process, where the nascent population of relativistic electrons would lose their energy before having a chance to propagate downstream in the jet to regions from which the opacity dropped sufficiently for the synchrotron
emission to be observable. In states A and B the soft X-ray
spectrum peaks at around $\sim1$ keV, and the synchrotron emitting
electrons probably have Lorentz factors $\gamma \geq 100$ (e.g. Fender
et al. 1999b) and so the inverse Compton process would scatter the photons out
of the easily observable X-ray band (the increase in energy is
typically by a factor of order $\gamma^2$). Is the photon field
sufficient to cool these electrons? A detailed calculation of this is
beyond the scope of this paper, but a simple estimate of the energy density of the photon field 50 Schwarzschild radii (a somewhat arbitrary guess at where the jet might be launched) above a $10^{38}$ erg s$^{-1}$ inner disc region
is of order $10^{10}$ erg cm$^{-3}$ (assuming a black hole mass of 10M$_{\sun}$). In such a photon field the lifetime of an electron of initial Lorentz factor $\sim100$ to inverse Compton losses would be $\sim10^{-5}$ s, which is not long enough for the electrons to escape to regions of significantly lower photons densities. Note that the spectral steepening related to the inverse Compton cooling has never been observed yet; this could be related to the fact that the transition between jet and cooled states is so rapid that we don't have a chance to catch the onset of cooling.

\section{Discussion}
\label{sec:discussion}

In this section we discuss the results of our analysis. Our main
concern is with the radio oscillation events, however, the radio
emission (pre- and post-plateau flares and the quasi-continuous
plateau jets) observed during the plateaux states will also be briefly
discussed. In the following whenever we refer to the term `jet', this
includes the pre- and post-plateau flares, the quasi-continuous
plateau jets and the radio oscillation events, which we assume are all
related to an outflow.

\subsection{The disc--jet connection}
\label{sec:djcon}

The analysis of the complete set of overlapping radio and X-ray
observations of the black hole candidate GRS~1915+105, presented in
the previous sections, has unambiguously shown the intimate relation
between the radio oscillation events and the spectrally hard State C episodes in the X-ray time series. As already shown by Belloni et al. (2000a), State C is the state in which GRS~1915+105 is most similar to the canonical black hole Low/Hard state (although it is not completely identical). A relation between the Low/Hard state in black hole X-ray binaries and their radio emission (quasi-continuous jets) has already been established
previously (Fender 2001): in the Low/Hard X-ray states of at least
three systems (Cyg~X-1, 1E~1740.7-2942 and GRS~1758-258)
quasi-continuous radio jets are directly imaged, whereas a general
feature of all `jet-systems' in the Low/Hard state (see Fender 2001,
Meier 2001) seems to be a flat or inverted synchrotron spectrum
extending from radio up to near IR or even sometimes optical
wavelengths. GRS~1915+105 is no exception to this rule, as both the
flat spectrum (Fig.~\ref{fig:platosc}) and the plateau jets themselves
are (directly) observed (Dhawan et al. 2000) during the plateau
states.

During the hard state (C) the X-ray spectrum is dominated by a power
law component, which is widely accepted to originate from inverse
Compton processes in the inner geometrically thick accretion flow
(such as a corona and/or an ADAF), although there are alternatives
(e.g. Markoff et al. 2001).  The empirical link between the
radio emission and these states, especially the one-to-one relation
(see Section~\ref{sec:onetoone}), strongly suggests that besides the outer
geometrically thin, and inner geometrically thick inflow there is an
outflow in the form of a jet. Fender et al. (1999a) have suggested
that during the Low/Hard state in black hole systems the Comptonizing
corona and the radio jet are strongly coupled. Also, based on a study
of the spectral and the timing properties of GRS~1915+105 during the
plateau states and the radio quiet states ($\chi_2$), Muno et al. (2001) come to
the conclusion that a coupling between the corona and the jet seems
very likely.

Assuming the jet and the corona are coupled, an injection of material
into the jet from the corona (or generation of a shock -- in the future we
will not repeat this alternative but consider it implicit) takes place
during State C intervals. If this ejection occurred at only one
point, for instance as the source moves in or out of State C, we would
expect the rise time of the radio emission to be shorter compared to
the case when the ejection takes place over a finite interval in
time. In fact from the arguments given in Section~\ref{sec:delays}, the
finite rise time indicates that the particle injection cannot be a
`delta function' but must also have a finite, observable duration.
From the observations shown in the previous sections we find the
rise time of the radio to be broadly comparable to the length of the State C
intervals ($\sim10$ -- $15$min), indicating that a more or less continuous ejection takes place during State C.

From the considerations given above the following scenario emerges for
the production of the radio oscillation events (see also sequence
shown in Fig.~\ref{fig:jets}, panel 3). When the source moves into
State C the thin disc truncates, as the inner radius moves outward (in the case of Fig.~\ref{abc50381}a, c and d, this is occurs within $3$ -- $5$ min). In
State C there is a large corona, and coupled to it a jet. The
outflow through the jet comes about through the more or less
continuous ejection of material from the corona. As a direct
consequence of this, longer State Cs are also radio brighter as more
material is being ejected. As the source moves out of State C the
accretion disc moves back in, a process which occurs on a longer
timescale ($\sim9$ -- $25$ min, see Fig.~\ref{abc50381}a, c and d),
eventually causing the injection of relativistic particles to
cease. The current radio data are not really of high enough quality to
see if there are measurable changes in the injection rate (ie. the
data seem to be consistent with a more or less constant gradient in
the radio light curve). Higher quality data may also be able to detect the
effects of the decay phase emission from both approaching and receding
components (assuming the ejections are symmetric), which will have
different Doppler factors. The result of the X-ray state change is
that we observe a discrete volume of synchrotron-emitting particles
moving downstream in the jet.

For sequences of repeating long ($\ga100$s) State C intervals we only observe the radio oscillation events when the separation between the State C intervals is large enough for the individual events to be observed: if the time it takes for the material to move up into the jet (from optically thick to thin) is longer than the recurrence time of the State C intervals, we can not distinguish between the individual events. The length of the State C intervals is not important in that respect: the length only determines the strength of the radio emission.

\subsection{Long and short State C intervals}
\label{sec:longshort}
In light of the scenario for the production of the radio oscillation events, suggested in Section~\ref{sec:djcon}, the question arises why series of short State C intervals do not also produce (high flux) radio oscillation events?

In Section~\ref{sec:statecinter} we have shown that for the (high flux) radio oscillation events to be observed, two conditions have to be met. The short State C intervals are not only too short ($\la100$s) to produce a high level of radio emission, but also the separation between the State C intervals is not enough for the individual events to be observed. 

To explain the disproportional high radio flux for the long State C intervals compared to the short State C intervals, we suggest that cooling of the material in the jet by soft X-rays from the State A and/or B (which are found in between the State C intervals; Section~\ref{sec:delays}) might be important. In the case of series of short State C intervals the soft photons of the disc can catch up with the material in the jet before it reaches the $\tau=1$ surface, as the State Cs are both short and close together, and hence the radio emission is suppressed.

\subsection{The low-level radio emission}
\label{sec:shortac}
It has become clear that State A, B and short State C intervals are related to a low-level radio emission (see Fig~\ref{abc50381}). The short State C intervals have already been discussed in the previous section. But why do we see radio emission in States A and B when there is no hard State C present (like in the case of class $\gamma$ and $\delta$)? 

One possibility is that the low-level radio emission we observe is indeed related to the decay of a large event occur before the interval in question, as noted in Section~\ref{sec:emp}. Another possibility is to realize that the absence of State C in the X-ray light curve does not necessarily mean that there is no hard component in the X-ray spectrum. In fact, as already mentioned in Section~\ref{sec:xrayvar} the power law component never seems to disappear completely during the soft States A and B. This weak, hard component could be associated with the ejection of a small amount of material during State A and B, which could explain the low level radio emission (also taking into account the possible cooling by soft photons as explained above). In that case the hardness of the spectrum is directly proportional with the strength of the radio emission. Although this is also suggested by Meier, Koide \& Uchida 2001 (see Section~\ref{sec:mhd}), we do not find a significant relation between the radio flux and the power law index, or between the radio flux and the size of the corona (estimated by the inner disc radius).

\subsection{Plateau states: the jets and flares}
\label{sec:jets}
In Section~\ref{sec:djcon} we discussed the relation between the hard State C intervals and the radio oscillation events. We can also extend this scenario to describe the presence of the quasi-continuous plateau jets, using the argument of continuous ejection during State C. In principle the plateau states are just larger, more stable versions of the State C intervals, only now a flat spectrum, quasi-continuous jet is observed as the jet is being fed from the corona for a longer period of time. This allows it to expand further out and hence become optically thin (see also Fig.~\ref{fig:jets}). The combined optically thick (at the base) and thin (further out) regions could produce the flat or slightly inverted radio spectrum, as already explained in Section~\ref{sec:platstates}.

In Fig.~\ref{fig:platosc} we show the radio -- infrared spectra for the plateau jets and the oscillation events. The similarity in the overall shape of the two spectra infers a possible common physical origin.

Two things remain unclear: why does the radio emission in class $\chi$ seems to saturate at a level of $\sim100$ mJy, and why is the radio emission lower during class $\chi_2$ compared to the plateau states? As far as the lower radio emission is concerned, as explained in Section~\ref{sec:radiobehaviour} $\chi_2$ seems to be different from $\chi_1$ and $\chi_3$ in several ways. Although more analysis is certainly necessary, the lower X-ray count rate during $\chi_2$ might reflect an insufficient supply of material that can be injected into the jet. With respect to the saturation level, an interaction between the continuous ejection and the possible cooling by soft photons might be responsible.

Summarizing, we have encountered three types of ejection events (see also Fig.~\ref{fig:jets}): 
\begin{itemize}
\item{the radio oscillation events, the plateau jets (class $\chi$) and the low-level radio emission during short State Cs, which seem to be related to the same type of mechanism operating on different scales;}
\item{the low-level radio emission during State A and B intervals, which can be related to a hard component but could also be a relic from a previous large radio event;}
\item{the pre- and post-plateau flares.}
\end{itemize}
The pre- and post-plateau flares differ from the other two types of ejection events in several ways. First of all the plateau flares are short discrete ejection events which become optically thin very rapidly and, in the case of $\chi_3$ are tracked to a distance of $\ga3000$ AU from the core (Mirabel \& Rodr\'\i guez 1994, Fender et al. 1999b, Dhawan et al. 2000). Secondly, contrary to the low-level radio emission, the radio oscillation events and the plateau jets (which are assumed to be related to quasi-continuous ejections) the plateau flares show clear indications of superluminal motion as separate blobs are observed to be ejected. Finally the pre- and post-plateau flares reach high flux levels up to $\sim$600 mJy (Fender et al. 1999b), on average a factor $\sim$3 -- 6 higher compared to the plateau jets and the radio oscillation events.

Based on the differences between the plateau flares and the other types of radio emissions, we speculate below that the plateau flares originate from a different mechanism.

\subsubsection{Plateau flares; a comparison with state transitions in other sources}

In the persistent black hole candidates Cyg~X-1 and GX~339$-$4 (Hjellming \& Han 1995, Brocksopp et al. 1999, Corbel et al. 2000) and in many Soft X-ray Transient Systems (SXTs) optically thin radio flares have been associated with state transitions (Fender \& Kuulkers 2001). In the case of the SXTs the flares are found during the fast transition from the `Off' (quiescence) state to the Soft state, but never in the opposite direction (note that the transition from Soft to Off in SXTs is usually much slower). For Cyg~X-1 and possibly also for GX~339$-$4 the flares have been associated with both the Soft -- Hard and the Hard -- Soft transitions (Zhang et al. 1997). The pre-plateau and the post-plateau flares, observed in GRS~1915+105 could be very closely related to the flares seen in Cyg~X-1, GX~339$-$4 and the SXTs: they occur as the source makes the transition into, or out of the plateau state which relates them to Soft -- Hard and Hard -- Soft transitions. 

The relation between the state transitions and the optically thin radio flares seems to be a general feature for the black hole systems. If the pre- and post-plateau flares are also related to state transitions, as argued above, GRS~1915+105 is one of the few systems to show the flares related to both the Soft -- Hard and Hard -- Soft transitions (note that this indirectly relates these flares to the hard State C intervals). In other systems the coverage is often poor and only occasionally (as in the case of the SXTs) flares are observed during the state transitions. 

\subsection{Xray~:~radio correlations during the $\chi$-states; a comparison to the Low/Hard states in GX~339$-$4 and Cyg~X-1}

In two other black hole jet sources, GX~339$-$4 and Cyg~X-1, there seems to be a strong three-way correlation between the radio, soft and hard X-ray emission during the Low/Hard state (Corbel et al. 2000, Brocksopp et al. 1999). During the High/Soft states in these two sources, both the radio emission and the hard X-rays are low. In the Low/Hard state GX~339$-$4 and Cyg~X-1 have a radio spectrum which is slightly inverted or flat ($\alpha$$\sim$$0.14$ and $\sim0$ respectively; see Fender 2001 and references therein), and during Low/Hard -- High/Soft state transitions the sources probably show discrete ejections of expanding relativistic plasma (Corbel et al. 2000, Brocksopp et al. 1999).

With respect to the Xray~:~radio coupling as presented in Section~\ref{sec:platstates}, it is clear that during class $\chi_2$ GRS~1915+105 resembles both GX~339$-$4 and Cyg~X-1: in all these sources there is a correlation between the radio and the soft X-ray emission. During $\chi_1$ and $\chi_3$, however, GRS~1915+105 shows an \emph{anticorrelation} instead of a correlation (Table~\ref{sec:x-ray:radio}). As far as the radio behaviour is concerned the jets and flares associated to $\chi_1$ and $\chi_3$ are most likely also found in GX~339$-$4 and Cyg~X-1. Although $\chi_2$ shows clearly a different radio behaviour, with no large radio jets or flares ever observed, the fact remains that the radio spectrum is flat (see Fig~\ref{fig:platosc}; a flat radio spectrum is considered to be the signature of a quasi-continuous jet) as also found for $\chi_1$, $\chi_3$, GX~339$-$4 and Cyg~X-1.

Concluding we find that, although there are similarities, GRS~1915+105 during class $\chi$ does not completely match the behaviour of normal black hole sources in the Low/Hard states. 

\subsection{Magneto-hydrodynamic jet formation}
\label{sec:mhd}

The formation of the jets is a process which is probably related to magnetohydrodynamic (MHD) effects (see Meier et al. (2001) for a review). In that case, the jets are produced/confined by the toroidal coiling of the magnetic field lines (which are embedded in the accretion disc) caused by the differential rotation of the accreting material (Meier et al. 2001). As stronger magnetic fields are expected for geometrically thick accretion flows, the appearance of a jet is directly linked to the Low/Hard states of black hole and neutron star X-ray binary systems (geometrically thick flows such as an ADAF or a corona are believed to be responsible for the production of the hard component in the spectrum). In fact, Meier (2001) shows that for X-ray binaries the most powerful jets are only produced in geometrically thick inner accretion flows where strong vertical (poloidal) magnetic fields are induced which are able to produce powerful ($\sim$10$^{38}$ erg~s$^{-1}$) radio jets. During the High/Soft states, on the other hand, when the spectrum is dominated by the soft multi-temperature black body component of the geometrically thin disc, the power in the jet is predicted to be a hundred times weaker compared to the thick flow solution (Meier 2001). This is in agreement with both our results and the observed relation between the black hole state and the jet in GX~339$-$4 by Fender et al. (1999a), and it could explain why during the soft States A and B only low-level radio emission is observed.

Although other processes such as shock fronts forming in the jets have been suggested and MHD processes still have difficulty explaining some of the high outflow speeds in the jet systems (see Meier et al. 2001) we feel, based on the arguments given above, that MHD processes are a good candidate for explaining the formation of the jets in the X-ray binary systems. Note however, that Meier (2001) and Meier et al. (2001) relate the strength of the radio emission to the hardness of the X-ray spectrum, while in our simple scenario (Section~\ref{sec:djcon}) it is related to the length of the State C intervals. 

\subsection{Alternatives}

In papers by Naik et al. (1999), Naik \& Rao (2000), Eikenberry et al. (2000) and Yadav (2001) it is concluded that that soft dips are related to the radio oscillation events. The observations presented in the papers can be classified as class $\beta$, $\lambda$, $\theta$ or $\nu$. We therefore suggest that in these observations the long ($\ga100$s), \emph{hard} State C dips, which are \emph{always} present just prior to State A dips (in the case when both States A and C are present; Belloni et al. 2000a), are responsible for the production of the radio oscillation events as explained in the previous sections. The fact that we can come to this conclusion is related to our much larger X-ray -- radio overlap compared to that of the authors mentioned above.

Naik et al. (1999) and Naik \& Rao (2000) also suggest that the combination of several soft dips produces the large pre- and post-plateau flares. Although we recognize that several small ejections can be combined to give one big radio event (such as a flare), we do not believe that is the case here. Probably the most convincing argument against this is the fact that we see several observations with repeating hard X-ray dips (eg. Figs.~\ref{abc50381}a,\ref{abc50698}c and \ref{abc51342}d) or with repeating soft dips (eg. Fig.~\ref{abc51432}e, and \ref{abc51342}d were soft State A dips are preceded by State C dips), during which no large plateau flares are reported (note that the argument presented here does not depend on whether the dips are hard or soft: in the model of Naik et al. (1999) and Naik \& Rao (2000) the soft dips are responsible for the ejection of material, in our model the hard dips are responsible). Moreover, Pooley \& Fender (1997) and Fender \& Pooley (1998, 2000) report multiple occasions of long sequences of radio oscillation events with \emph{no} major flare. Finally, a close examination of similar outbursts to the one discussed by Naik et al. (1999) reveals that the first series of long State C dips (observations K-52-01 and K-52-02, see Table~\ref{t1}), and therefore also the corresponding radio oscillation events, start \emph{after} the large jet is observed (see figure~4 in Fender et al. 1999b). Taking into account the delay between the X-rays and the radio emission, it is very unlikely that the combined effect of these small ejections could be responsible for the production of the large plateau jets.

In two different papers Nayakshin, Rappaport, Melia (2000) and Janiuk, Czerny, Siemiginowska (2000) conclude they need an outflow in the form of a jet, to correctly model the complex behaviour of GRS~1915+105. They both model the jet outflow in the same way assuming the ejections set in as the source makes the transition into the High/Soft state. As explained in the previous sections, this is inconsistent with our results: we suggested, based observational facts, that the ejection of material does not occur at a single point but is more or less continuous during State C. Therefore we feel that future models should incorporate jet outflows which are related to hard (State C) intervals which could, in principle, lead to the production of the radio oscillation events and the plateau jet in a natural way (as explained in Section~\ref{sec:djcon}).

\section{Summary \& Conclusions}
\label{sec:conclusions}

\begin{table}
\caption{The emipircal X-ray~:~radio correlation in GRS~1915+105.\label{t4}}
\begin{tabular}{cc}
\hline
Radio Behaviour	&X-ray behaviour\\
\hline
\hline
Radio oscillation events &repeating, hard, well\\
 &separated $\ga100$s State C\\
\hline
Low level emission & adjacent, $\la100$s \\
 &State C or A, B\\
\hline
Plateau Jet &long uninterrupted\\
 &State C\\
\hline
Pre-, Post-plateau flares &State transitions?\\
\hline
\end{tabular}
\end{table}

This work has clearly established that radio emission from GRS~1915+105 is intimately related to the presence of hard (power-law-dominated) intervals in the X-ray light curves. This in turn physically implies a clear relation between a radiatively inefficient flow close to the black hole, and a synchrotron-emitting outflow or jet. 

In Table~\ref{t4} we summerize the empirical X-ray~:~radio correlation, moreover we find that:
\begin{itemize}
\item{The (high flux) radio oscillation events are closely related to a series of long (or low count rate), well-separated, spectrally hard State C intervals in the X-ray light curve.}
\item{The State C intervals must be at least 100s in length and have a recurrence time that allows for the individual radio oscillation events to be observed.}
\item{We confirm a one-to-one relation between radio oscillation events and series of long State C intervals: each State C interval produces a new ``radio flare''.}
\item{Short ($\la100$s) State C intervals and State A and B intervals produce a similar low-level radio emission.}
\item{The radio oscillation profiles seem to be determined by particle injection/acceleration (rise phase) and adiabatic expansion losses (decay phase), which are both frequency independent effects.}
\item{During class $\chi_1$ and $\chi_3$ (plateau states) we find a quasi-continuous jet with a flat synchrotron spectrum (extending to at least the near-infrared) and a flux level of $\sim100$ mJy, an anticorrelation between the radio flux and the soft (ASM) X-ray flux, and a correlation between the radio flux and the ASM HR$_{2}$ colour.}
\item{During class $\chi_2$ we also find a flat synchrotron spectrum, but a flux level which is about a factor 10 lower, a correlation between the radio flux and the soft (ASM) X-ray flux and a anticorrelation between the radio flux and the ASM HR$_{2}$ colour.}
\end{itemize}

Based on this we suggest a simple scenario which describes the complex X-ray~:~radio behaviour of GRS~1915+105:
\begin{itemize}
\item{During State C a more or less continuous ejection of relativistic particles takes place;}
\item{The length (or depth, as longer also means deeper) of the State C interval determines the strength of the radio emission: longer State Cs have a higher radio flux;}
\item{The separation between the State Cs determines the shape of the radio light curve: if the intervals are to close together one observes a ``stable'' radio flux level, while if the separation is large enough for the individual radio flares to be observed one finds a radio light curve dominated by oscillations.}
\end{itemize}

Within the simple scenario we can explain:
\begin{itemize}
\item{the radio oscillation events: series of well separated, long State Cs;}
\item{the radio plateau jets: uninterrupted State Cs;}
\item{the low-level radio emission during the short State Cs: short, near-contiguous State Cs.}
\end{itemize}
All these types of radio emission seem to be related to the same physical mechanism operating on different scales.

However, this scenario does not explain:
\begin{itemize}
\item{The disproportional higher radio flux in the case of the long State C compared to the short State C intervals (although cooling by soft X-ray photons might be responsible);}
\item{The low-level radio emission in the case of State A and B, which cannot be explained based on argument of a continuous ejection during State C;} 
\item{The radio plateau flares observed prior to and just after the plateau states, which seem to be related to the Soft -- Hard and Hard -- Soft X-ray state transitions as also observed in SXTs;}
\item{The lower radio emission during class $\chi_2$ compared to classes $\chi_1$ and $\chi_3$.}
\end{itemize}

With respect to the low-level radio emission found during States A, B and short State C intervals, we cannot rule out the possibility that the emission is a relic from a previous large radio event. 

The most important result is probably the establishment of the State C -- radio relation, and in particular the suggested one-to-one relation between the long State C intervals and the radio oscillation events. Because GRS~1915+105 has a much larger coverage in the radio than in the X-rays (PCA observations), this intimate relation can provide us with much more information about the behaviour of the source its behaviour than is revealed to us by the X-rays alone. However, due to the time delay between the radio and the X-rays, the disc -- jet connection is probably better studied in the X-ray and IR bands. 

The analysis presented here, together with previous reports on other black hole X-ray binary systems such as GX~339$-$4 (Fender et al. 1999a) and XTE~J1550-564 (Corbel et al. 2001), provides strong evidence that hard, power law-dominated states can support a radio-emitting outflow. This suggest, as discussed in Section~\ref{sec:mhd}, that MHD effects could be responsible for the production and/or confinement of the jets found in these systems.

\section*{Acknowledgements}

The authors would like to thank Jeroen Homan for carefully reading this document, and for providing useful comments and suggestions. This research has made use of data obtained through the High Energy Astrophysics Science Archive Research Center Online Service, provided by the NASA/Goddard Space Flight Center. We acknowledge the use of quick-look results provided by the ASM/\emph{RXTE} team, and thank the staff at MRAO for maintenance and operation of the Ryle Telescope, which is supported by the PPARC. 

{}

\end{document}